\documentclass[twocolumn,eqsecnum,aps,showpacs,prb,superscriptaddress]{revtex4}
\usepackage{graphicx}
\usepackage{dcolumn}
\usepackage{amsmath,amssymb}
\usepackage{cancel}
\usepackage{makeidx}
\usepackage{amsfonts}

\begin{document}

\title{On calculating the Berry curvature of Bloch electrons using the KKR
method}
\author{M.~Gradhand}
\email{martin.gradhand@physik.uni-halle.de}
\affiliation{Max-Planck-Institut f\"ur Mikrostrukturphysik, Weinberg 2, D-06120 Halle,
Germany}
\author{D.V.~Fedorov}
\affiliation{Institut f\"ur Physik, Martin-Luther-Universit\"at Halle-Wittenberg, D-06099
Halle, Germany}
\author{F.~Pientka}
\affiliation{Institut f\"ur Physik, Martin-Luther-Universit\"at Halle-Wittenberg, D-06099
Halle, Germany}
\affiliation{Dahlem Center for Complex Quantum Systems and Fachbereich Physik, Freie Universit\"at Berlin, D-14195 Berlin, Germany}
\author{P.~Zahn}
\affiliation{Institut f\"ur Physik, Martin-Luther-Universit\"at Halle-Wittenberg, D-06099
Halle, Germany}
\author{I.~Mertig}
\affiliation{Institut f\"ur Physik, Martin-Luther-Universit\"at Halle-Wittenberg, D-06099
Halle, Germany}
\affiliation{Max-Planck-Institut f\"ur Mikrostrukturphysik, Weinberg 2, D-06120 Halle,
Germany}
\author{B.L.~Gy\"orffy}
\affiliation{H.H.Wills Physics Laboratory, University of Bristol, Bristol BS8 1TH, United Kingdom}

\begin{abstract}
We propose and implemented a particularly effective method for calculating 
the Berry curvature arising from adiabatic 
evolution of Bloch states in ${\bf k}$ space. The method exploits 
a unique feature of the Korringa-Kohn-Rostoker (KKR) approach to solve 
the Schr\"odinger or Dirac equations. Namely, it is based on the observation 
that in the KKR theory the wave vector ${\bf k}$ enters the calculation only via the
structure constants which reflect the geometry of the lattice
but not the crystal potential. For both the Abelian and non-Abelian Berry curvature we derive an
analytic formula whose evaluation does not require any numerical differentiation
with respect to ${\bf k}$. We present explicit calculations for Al, Cu, Au, and Pt bulk crystals.
\end{abstract}

\pacs{71.15.-m, 71.15.Rf, 71.15.Dx}
\maketitle

\section{Introduction}

Over the past decade it has been realized that the Berry curvature 
$\boldsymbol\Omega_n ({\bf k})$ associated with Bloch waves 
in solids can play an important role in spin and charge transport 
by electrons.\cite{Berry84,Bohm03} Consequently, a first principle calculation of this 
qunatity was highly desirable. Important examples where such calculations 
have already been found useful are the Anomalous Hall Effect (AHE) 
\cite{Yao04,Wang06,Wang07} and the Spin Hall Effect (SHE) \cite{Yao05,Guo05,Guo08}. 
Particularly insightful are those which focus on the integral of 
$\boldsymbol\Omega_n ({\bf k})$ over the Fermi surface only. Following Haldane's
suggestion,\cite{Haldane04} it was applied in Ref.~\onlinecite{Wang07}.

The methodologies used in these calculations are based on two disctinct approaches. 
One is the evaluation of the Kubo formula for the off-diagonal elements $\sigma _{xy}$ 
of the static conductivity.\cite{Thouless82,Yao04,Yao05,Guo05,Guo08} The other one uses 
the first principles Wannier representation of the Bloch states.\cite{Wang06,Wang07} In what follows, we present an alternative way constructed 
within the framework of the KKR approach.\cite{Korringa47,Kohn54}

Since the most interesting problems, where the above
curvature is relevant, concern the role of spin-orbit coupling, we want to
develop our approach for a fully relativistic description of the electronic
structure. To be more specific we recall that for the Dirac Bloch wave of
the conventional form\cite{Strange98}
\begin{equation}
\begin{array}{ll}
\Psi_{n {\bf k}}({\bf r})=e^{i {\bf k}\cdot {\bf r}} 
u_{n {\bf k}}({\bf r})\ ,  
\end{array}
\label{eqno(1)}
\end{equation} 
the \emph{connection} corresponding to the \emph{geometrical phase} $\gamma_n({\bf k})$ is defined as
\begin{equation}
\begin{array}{ll}
{\bf \cal A}_{n}({\bf k}) = i\int\limits_\omega u_{n {\bf k}}^\dagger ({\bf r})
\boldsymbol\nabla_{\bf k} u_{n {\bf k}} ({\bf r}) d {\bf r}\ ,
\end{array}
\label{eqno(2)}
\end{equation}
where the integral is over a unit cell of the volume $\omega$. Then, the corresponding \emph{curvature} 
is given by
\begin{equation}
\begin{array}{ll}
\boldsymbol\Omega_n ({\bf k}) = \boldsymbol\nabla_{\bf k} \times
{\bf \cal A}_{n}({\bf k})\ .
\end{array}
\label{eqno(3)}
\end{equation}
In the above notation $n$ is a band index and $u_{n {\bf k}}({\bf r})$ is a periodic four 
component spinor function of ${\bf r}$.

Here we shall demonstrate that KKR-based band theory methods 
are particularly well suited for the task. Our central point is that
the KKR matrix, whose determinant is conventionally used to find the energy
bands, has its own well defined \emph{geometrical phases, connections} and
\emph{curvatures}. As well as being easy to calculate, they are
closely related to those defined above. The root cause of this convenient
feature is the fact that such matrices depend parametrically on the wave
vector ${\bf k}$ and the energy ${\cal E}$. Therefore, the geometry of their 
eigenvalues and eigenvectors, in the ${\bf k}$ and ${\cal E}$ space, 
is closely related to that associated with the periodic part of the Bloch 
functions.\cite{Bohm03} There are three factors 
which make the study of KKR matrices computationally efficient. Firstly, by 
the standards of first principles electronic structure calculations the 
ranks of KKR matrices are quite small. Typically, one is dealing with 16$\times
16$ (Schr\"odinger equation) or $32\times 32$ (Dirac equation) matrices
(if we assume one atom per unit cell). Secondly, the crystal momentum
${\bf k}$ enters into the computation only through the structure constants 
and their gradients with respect to ${\bf k}$. Furthermore, these quantities 
depend only on the geometrical crystal structure but not the crystal potential. 
So, they are readily calculated, without taking numerical derivatives, 
by using the so-called screened version of the KKR method.\cite{ZahnPhD,Zeller95} 
Finally, the calculations can proceed in the constant energy mode which is particularly efficient when studying Fermi surface properties.

To demonstrate the efficiency and stability of the proposed numerical procedures, 
we present explicit calculations of the Berry curvature on the Fermi 
surfaces of Al, Cu, Au, and Pt bulk crystals. In a fully relativistic theory the presence of both space and time
inversion symmetry forces every ${\bf k}$ state to be
twofold degenerate.\cite{Elliott54,Kramers30} As a consequence, we have to deal
with the so-called \emph{non-Abelian} Berry curvature.\cite{Wilczek84,Shindou05} The corresponding
formalism is derived within this paper. In order to illustrate the significance of such calculations 
for the understanding of interesting physical phenomena, we computed the intrinsic contribution 
to the spin Hall conductivity for Pt and Au. We compare our results with those 
obtained by other methods.\cite{Yao05,Guo08APL} Although these calculations were performed 
using the usual Fermi sea integration, our formalism should be especially efficient for
 approaches based on Haldane's suggestion.\cite{Haldane04} According to Haldane the calculations require 
the Berry curvature only on the Fermi surface.

We will introduce our novel theoretical framework for calculating the above
\emph{connection} and the \emph{curvature} in two steps. In Section II we 
present an alternative for computing the group velocity
\begin{equation}\label{Eq.:velocity}
{\bf v}_n ({\bf k}) = \boldsymbol\nabla_{\bf k} 
{\cal E}_n ({\bf k})
\end{equation}
of Bloch electrons, without taking the partial derivative with respect 
to ${\bf k}$ numerically. In Section III we extend this approach to 
the calculation of the Berry curvature in both, Abelian and non-Abelian, cases. 
The example computations are shown and discussed in Section IV. 
Note that in Sections I-IV we use atomic units with energy in Rydberg. 
The results for the spin Hall conductivity are presented in Section V. We conclude in Section VI
and the Appendix provides a detailed derivation of the formulas used in the calculations.

\section{Calculating the Group velocity for Bloch electrons}

In this section we prepare the ground for our principle task in Section III by outlining a simple, 
instructive way of computing the group velocity ${\bf v}_n ({\bf k})$. In addition, we introduce briefly 
the relativistic KKR formalism.\cite{Zabloudil2005,Gradhand09} The basic 
idea for calculating the group velocity was suggested by Shilkova and Shirokovskii in Ref.~\onlinecite{Shilkova88}.  
However, our procedure will follow a slightly different route 
and hence will be described in detail below.

To be specific with regard to notation we use that of Ref.~\onlinecite{Gradhand09}.
Here we restrict our consideration to the non spin-polarized case that means
nonmagnetic systems. To simplify the equations, we assume
one atom per unit cell. Nevertheless, the generalization to a lattice with a basis
is straightforward. In addition, we use the atomic-sphere approximation (ASA) for the
crystal potential in the Dirac equation.\cite{Gradhand09}

Then the Bloch wave corresponding to a band $n$ can be expanded around a site in the ASA sphere as
\begin{equation}
\begin{array}{ll}
\Psi_{n {\bf k}} ({\bf r}) = \sum\limits_Q C_Q^n ({\bf k})
\Phi_Q ({\cal E}_n({\bf k});{\bf r})\ ,
\end{array}
\label{KKRexpansion}
\end{equation}
where
\begin{equation}
\begin{array}{ll}
\Phi_Q ({\cal E};{\bf r}) =
\left( \begin{array}{c}
g_\varkappa ({\cal E};r) \chi_Q ({\bf e}_r) \\ i f_\varkappa ({\cal E};r) 
\chi_{\bar Q}({\bf e}_r) \end{array} \right)
\end{array}
\label{basis_set}
\end{equation}
are the scattering solutions of the Dirac equation 
for the spherically symmetric potential at the energy ${\cal E}$. They are written
in terms of the large and the small component, where $g_\varkappa ({\cal E};r)$ and $f_\varkappa ({\cal E};r)$ are
the corresponding radial functions.\cite{Zabloudil2005,Gradhand09} Here $Q = \{\varkappa ,\mu \}$ and ${\bar Q}= \{-\varkappa ,\mu \}$ are 
abbreviations for the quantum numbers $\varkappa$ and $\mu$
specifying the conventional spin-angular eigenfunctions $\chi_Q({\bf e}_r)$,\cite{Strange98} where ${\bf e}_r = {\bf r}/r$. 

When the multiple scattering ideas of Korringa \cite{Korringa47} and Kohn and Rostoker 
\cite{Kohn54} are invoked, one finds that the energy eigenvalues ${\cal E}_n ({\bf k})$ 
are given by those combinations of ${\cal E}$ and ${\bf k}$ for which the determinant 
of the KKR matrix
\begin{equation}
\begin{array}{ll}
M_{Q Q^\prime} ({\cal E}; {\bf k}) = G_{Q Q^\prime}^{s} ({\cal E}; {\bf k})
\Delta t_{Q^\prime}^{s} ({\cal E})- \delta_{Q Q^\prime}  
\end{array}
\label{KKR-matrix}
\end{equation}
is zero. Note that the screened structure constants G$_{Q Q^\prime}^{s} 
({\cal E}; {\bf k})$ \cite{Zeller95} depend only on the crystal structure 
while the screened \emph{$\Delta$t-matrix} 
describes the scattering at the local, self-consistent 
effective one-particle potential. Therefore, $\Delta t_{Q}^{s}({\cal E})$ is a function of energy ${\cal E}$ 
but not of ${\bf k}$. This is the separation of crystal structure 
and potential mentioned in the introduction. 
Moreover, the more sophisticated, and physically more relevant, spin-polarized version of the 
theory will retain the formal structure with the difference that
the \emph{$\Delta$t-matrix} will be non diagonal in Q.

An efficient way of finding the zeros of the KKR determinant 
$|| M_{Q Q^\prime} ({\cal E}; {\bf k}) || $ is to solve 
the matrix eigenvalue problem
\begin{equation}
\begin{array}{ll}
\Bar{\Bar M} ({\cal E}; {\bf k}) {\bar C}_n =
\lambda _{n} {\bar C}_n  
\end{array}
\label{eqno6}
\end{equation}
and to search for vanishing eigenvalues $\lambda_n ({\cal E}; {\bf k})$. It can be performed, 
either in ${\bf k}$ space at constant energy or in ${\cal E}$ at fixed ${\bf k}$.
In the above notation the components of the matrix $\Bar{\Bar M}$ and the $n$th eigenvector 
${\bar C}_n = \{ C_Q^n ({\bf k}) \}$ are labeled by $Q$. 

By means of the expansion coefficients ${\bar C}_n$, corresponding to the band 
energy ${\cal E}_n ({\bf k})$, we could calculate the group velocity evaluating
\begin{equation}
\begin{array}{ll}
{\bf v}_n ({\bf k}) = \int\limits_\omega 
\Psi_{n {\bf k}}^\dagger ({\bf r}) c
\boldsymbol{\hat \alpha} \Psi_{n {\bf k}} ({\bf r}) d {\bf r}\ ,
\end{array}
\label{eqno7}
\end{equation}
with the relativistic velocity operator $c 
\hat{\boldsymbol\alpha}$.
As was shown, analytically, by Shilkova and Shirokovskii \cite{Shilkova88}
this formula is equivalent to Eq. (\ref{Eq.:velocity}). However, within the ASA approximation used in this paper, 
the expression (which follows from Eqs.~(\ref{KKRexpansion}) and (\ref{eqno7}))  
\begin{equation}
\begin{array}{ll}
{\bf v}_n ({\bf k}) = {\bar C}_n^\dagger ({\bf k}) c \boldsymbol{\Bar{\Bar \alpha}} 
({\cal E}) {\bar C}_n ({\bf k})\ ,
\end{array}
\label{eqno8}
\end{equation}
where the elements of the vector matrix $\boldsymbol{\Bar{\Bar \alpha}}$ are defined as
\begin{equation}
\begin{array}{ll}
(\boldsymbol{\Bar{\Bar \alpha}})_{Q Q^\prime} ({\cal E}) \equiv
\int\limits_{\omega } \Phi_Q^\dagger ({\cal E}; {\bf r}) \boldsymbol{\hat \alpha}
\Phi_{Q^\prime} ({\cal E}; {\bf r}) d {\bf r}\ ,
\end{array}
\end{equation}
does not reproduce the results of the numerical differentiation exactly. 
We will comment on this problem at the end of the current section.

For now we turn to the central result of Ref.~\onlinecite{Shilkova88} which is based 
on Eqs.~(\ref{Eq.:velocity}), (\ref{eqno6}) and derive a similar expression. Technically, 
the solution would be easier if the matrix $\Bar{\Bar M} ({\cal E}; {\bf k})$ 
was Hermitian. However, due to the used expansion, it is not. An additional transformation, discussed by Kohn and Rostoker \cite{Kohn54} and 
used in our previous papers based on Refs.~\onlinecite{ZahnPhD} and \onlinecite{Gradhand09}, can provide a Hermitian 
KKR matrix.\cite{Comment2} However, the derivation of the Berry curvature described in Section III would 
be more complicated due to the necessary normalization of the basis functions.		 
Thus, for clarity, we proceed to solve Eq.~(\ref{eqno6}) for a non-Hermitian matrix 
$\Bar{\Bar M} ({\cal E}; {\bf k})$. 
In short, we find the right and left eigenvectors, 
${\bar C}_n ({\cal E}; {\bf k})$ and ${\bar D}_n ({\cal E}; {\bf k})$, respectively, 
such that the following conditions are fulfilled: ${\bar C}_n^\dagger {\bar C}_n =1$,
${\bar D}_n^\dagger {\bar D}_n =1$. We note that ${\bar D}_n^\dagger {\bar C}_{n^\prime} \propto
\delta_{n n^\prime}$, ${\bar C}_n^\dagger {\bar C}_{n^\prime} \ne 0$,
${\bar D}_n^\dagger {\bar D}_{n^\prime} \ne 0$, and ${\bar D}_n^\dagger {\bar C}_n \ne 1$.
Here ${\bar C}_n$ and ${\bar D}_n$ correspond to the same eigenvalue $\lambda _{n}$.
A straightforward algebra, summarized in Appendix \ref{App_a}, yields 
\begin{equation}
\begin{array}{ll}
{\bf v}_n ({\bf k}) = - \frac{D_n^\dagger (\frac{\partial \Bar{\Bar M} ({\cal E}; {\bf k})}
{\partial {\bf k}}\vert_{_{_{{\cal E}={\cal E}_n ({\bf k})}}}) C_n}
{(D_n^\dagger C_n) \frac{\partial\lambda_n ({\cal E})}{\partial {\cal E}}
\vert_{_{_{{\cal E}={\cal E}_n ({\bf k})}}}}\ .
\end{array}
\label{eqno8c}
\end{equation}

This expression, being the main result of the current section, is similar to the one obtained in 
Ref.~\onlinecite{Shilkova88} for the Hermitian KKR matrix. It shows that having found 
the Bloch state energy ${\cal E}_n ({\bf k})$, that provides the zero of the $n$th eigenvalue 
$\lambda_n ({\cal E}; {\bf k})$, one can calculate the velocity by evaluating the 
above formula. For this purpose, the eigenvectors ${\bar C}_n ({\cal E}; {\bf k})$ and 
${\bar D}_n ({\cal E}; {\bf k})$ corresponding to 
$\lambda_n ({\cal E}_n ({\bf k}); {\bf k})=0$ as well as the partial derivative 
of $\Bar{\Bar M} ({\cal E}; {\bf k})$ with respect to ${\bf k}$ are required. 
Since in the screened KKR method $\partial \Bar{\Bar M} / \partial {\bf k}$ can be evaluated analytically, 
the disadvantage of taking numerical derivatives of the dispersion relation ${\cal E}_n ({\bf k})$
is avoided. Namely, it follows from Eq.~(\ref{KKR-matrix}) that 
$\partial \Bar{\Bar M} ({\cal E}; {\bf k}) / \partial {\bf k} = (\partial \Bar{\Bar G}^s 
({\cal E}; {\bf k}) / \partial {\bf k}) \Delta t^s ({\cal E})$. Noting that, one can use 
the short range feature of the screened real space structure constants 
$\Bar{\Bar G}^s ({\cal E}; {\bf R})$\cite{Zeller95} to evaluate
\begin{equation}
\begin{array}{ll}
\partial \Bar{\Bar G}^s ({\cal E}; {\bf k}) / \partial {\bf k} =
i \sum\limits_{\bf R} {\bf R} e^{i {\bf k} {\bf R}} \Bar{\Bar G}^s ({\cal E}; {\bf R})
\end{array}
\label{eqno8d}
\end{equation}
at each ${\bf k}$ point, separately. Consequently, the only 
numerical derivative to be taken, by calculating the velocity, is the one-dimensional derivative 
$\partial \lambda_n ({\cal E}; {\bf k}) / \partial {\cal E}$. Fortunately, this requires 
only modest computational efforts.

In concluding this section, we report in Fig.~1 a comparison between 
$\boldsymbol\nabla_{\bf k} {\cal E}_n ({\bf k})$ as calculated by numerical differentiation, 
by the use of Eq.~(\ref{eqno8}), and by evaluating the formula of Eq.~(\ref{eqno8c}).  
The calculations are performed for the electron states on the Fermi surface of Cu. Significantly, the results based on Eqs.~(\ref{eqno8}) and (\ref{eqno8c}) show 
a smooth appearance over the Fermi surface, indicating their independence on the number of 
${\bf k}$-mesh points. By contrast, the numerical derivative in Eq.~(\ref{Eq.:velocity}) strongly depends on the used ${\bf k}$ mesh. 
The other noteworthy features of these results are the similarities and differences of the velocities obtained by Eqs.~(\ref{eqno8}) 
and (\ref{eqno8c}). A detailed analysis of those shows that 
the numerical derivative of $\boldsymbol\nabla_{\bf k} {\cal E}_n ({\bf k})$ converges to 
the result of Eq.~(\ref{eqno8c}) whereas for the direct evaluation of the velocity operator 
by Eq.~(\ref{eqno8}) a maximal error of $4\%$ remains. This effect was already discussed in 
the literature with respect to dipole transition matrix elements.\cite{Shilkova88_a,Guo_95} 
It was shown that the ASA approximation causes difficulties in evaluating the off-diagonal 
matrix elements of the relativistic velocity operator. The authors of Refs.~\onlinecite{Shilkova88_a} and \onlinecite{Guo_95} resolved the issue by rewriting 
the necessary formulas to get numerically more stable results. The problem in evaluating 
the expectation value of $c\hat{\boldsymbol \alpha}$ was already discussed by Shilkova and Shirokovskii who 
solved the problem by following the line of arguments we have adopted here. They showed that this method is perfectly stable. Here we confirm their results 
for the case of non-Hermitian KKR matrices. 

Finally, we point out that the method of calculating the Berry curvature presented in the next section uses the same 
techniques as considered above. Therefore, similar improvements of accuracy and stability for the numerical results are 
expected.
\newlength{\LL} \LL 0.32\linewidth
\begin{figure}[ht]
\includegraphics[width=\LL]{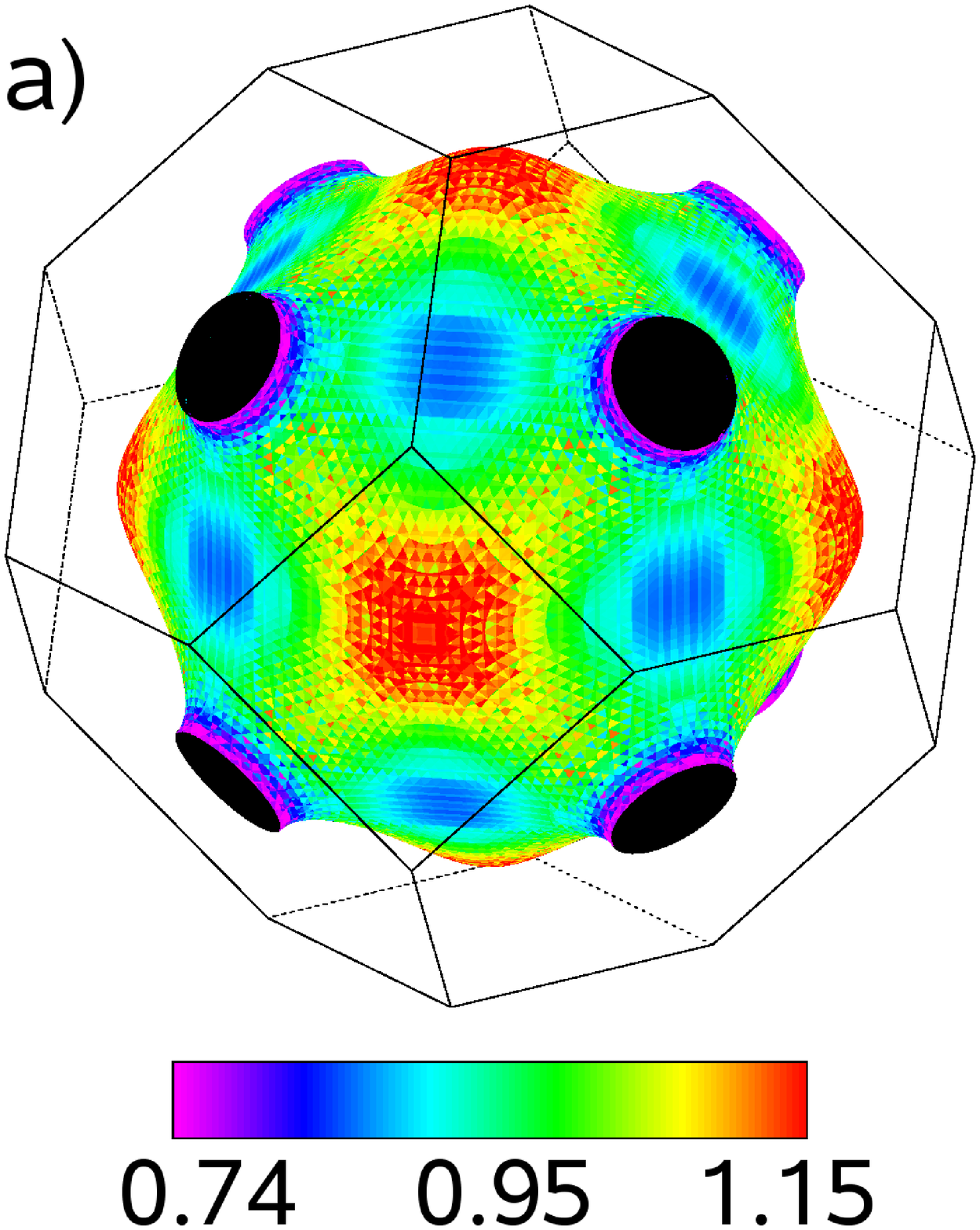}
\includegraphics[width=\LL]{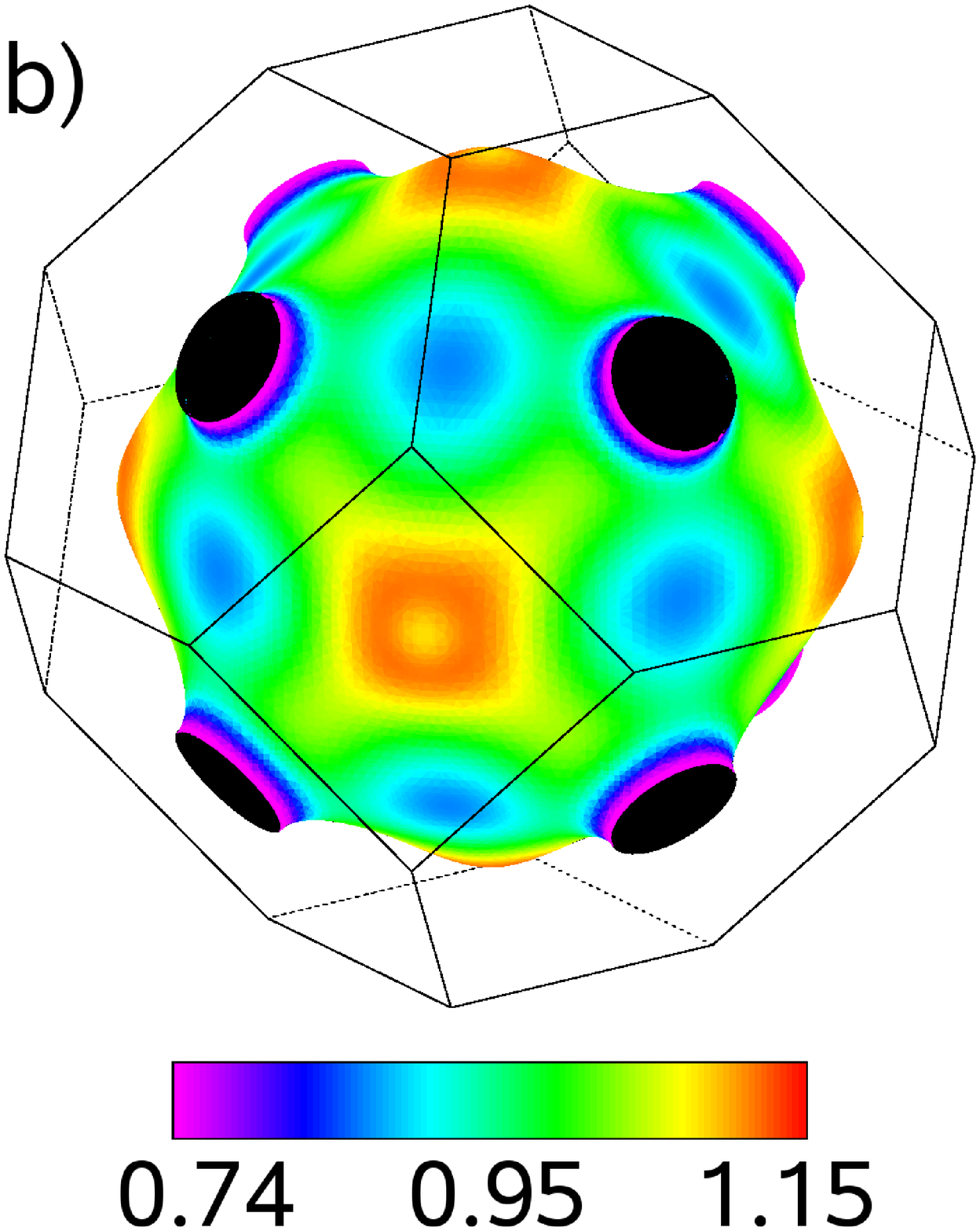}
\includegraphics[width=\LL]{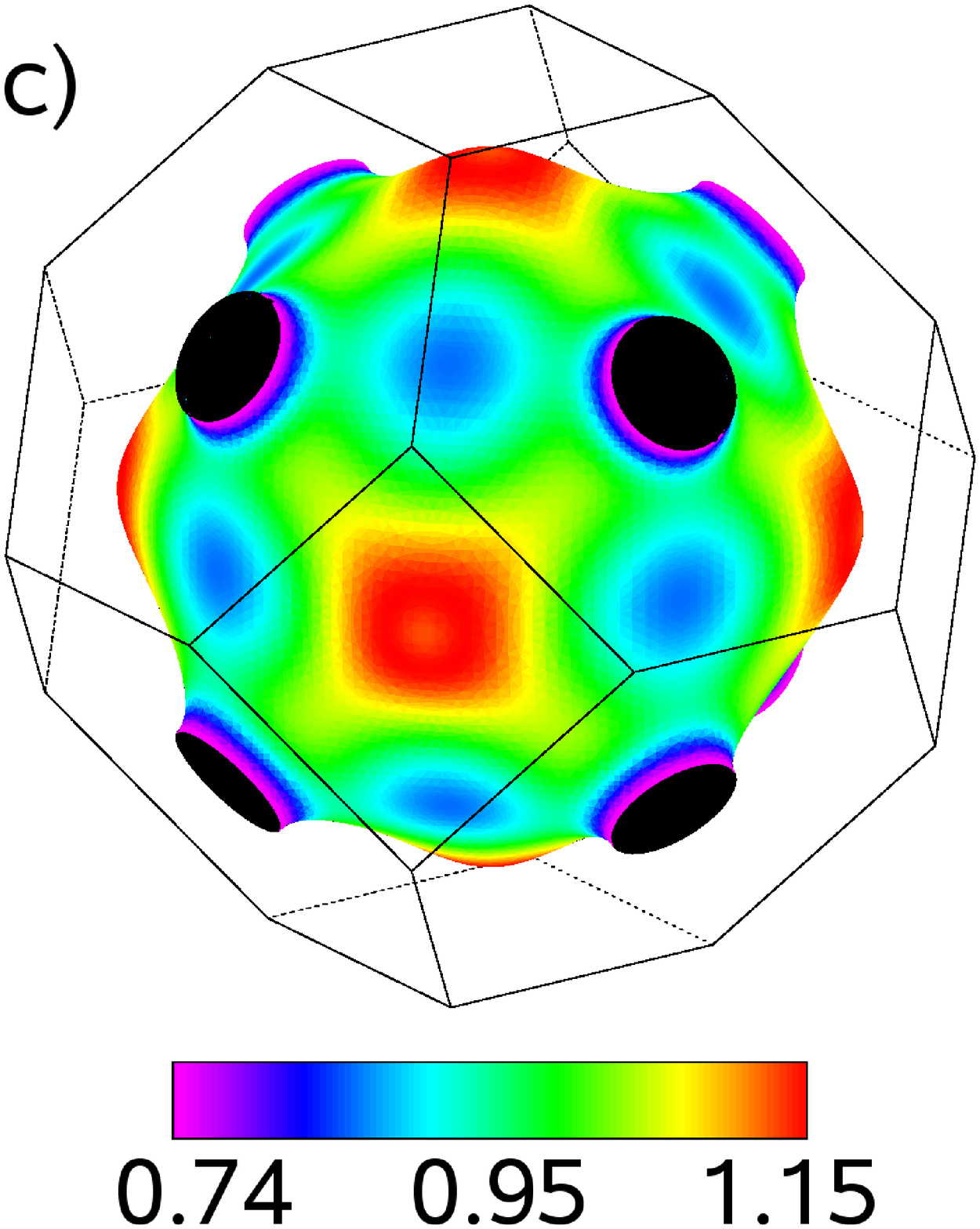}
\caption{The absolute value of the Fermi velocity of Cu (in a.u.) obtained using three different 
methods: a) the numerical derivative of the dispersion relation 
$\boldsymbol\nabla_{\bf k} {\cal E}_n ({\bf k})$; b) the expectation value of the relativistic 
velocity operator given by Eq. (\ref{eqno8}); and c) implementation of Eq. (\ref{eqno8c}) }
\label{fig.:v_num_alpha}
\end{figure}
 
\section{ New route to compute the Berry curvature}

In this section the formalism for the calculation of the Berry curvature within the KKR method 
is derived. We start with the conventional (Abelian) case for ${\bf \cal A}_{n}({\bf k})$ and 
$\boldsymbol\Omega_n ({\bf k})$ (subsection A and B, respectively). Then we expand our
consideration to a general non-Abelian case (subsection C).

\subsection{The \emph{connection} for $u_{n {\bf k}}({\bf r})$ via 
$\Psi_{n {\bf k}}({\bf r})$}

Clearly, the periodic part 
$u_{n {\bf k}}({\bf r})$ of the Bloch wave 
is an eigensolution of the Schr\"odinger or Dirac equation with Hamiltonian 
\begin{equation}
\begin{array}{ll}
{\hat H}_{\bf k} ({\bf r}) = e^{-i {\bf k}\cdot{\bf r}} {\hat H} ({\bf r}) 
e^{i {\bf k}\cdot{\bf r}}
\end{array}
\label{kHamiltonian}
\end{equation}
in which the wave vector appears as a parameter. Thus, the arguments 
leading to Eqs.~(\ref{eqno(2)}) and (\ref{eqno(3)}) are, by now, conventional.\cite{Bohm03} 
However, whether the Bloch wave 
itself has a \emph{geometrical phase}, \emph{connection} and \emph{curvature} 
in its own right appears to be a different problem. The Hamiltonian for $\Psi_{n {\bf k}}({\bf r})$ does
not depend on ${\bf k}$, and the wave vector enters into the discussion of Bloch
waves only by defining the boundary conditions. Although it has been noted \cite{Resta2000}, it was not clarified 
whether a slowly changing boundary condition is exactly equivalent (or has the same
holonomy) to a slowly changing parameter ${\bf k}$ in the theory of 
$u_{n {\bf k}}({\bf r})$. 

Another comment which concerns the above discussion is that ${\bf k}$ of $\Psi_{n {\bf k}} ({\bf r})$ 
labels not only the energy eigenstate but also the eigenvalues 
$e^{i {\bf k}\cdot {\bf R}}$ of the translation operators ${\hat T}_{\bf R}$. Therefore, 
it is not entirely free to act as a parameter. By contrast, $u_{n {\bf k}} ({\bf r})$ 
is degenerate with respect to all translation operators and hence its ${\bf k}$ is 
not obliged to label their eigenvalues. As a consequence, they are free to be parameters 
in ${\hat H}_{\bf k}({\bf r})$. In other words, $u_{n {\bf k}}$ is not in the same Hilbert 
space as $u_{n {\bf k}^\prime}$ and hence they do not need to be orthogonal. 
In contrast,  $u_{n {\bf k}}$ and  $u_{m {\bf k}}$ with $m\neq n$ reside in the same Hilbert space and are orthogonal to each other.\cite{Resta2000}
 
With these remarks in mind we note that the KKR, as most band-theory methods, is 
designed to calculate $\Psi_{n {\bf k}} ({\bf r})$ 
but not $u_{n {\bf k}} ({\bf r})$, in addition to the energy eigenvalue 
${\cal E}_n ({\bf k})$. Nevertheless, the Bloch function in the unit cell $\omega$, 
as given by Eq.~(\ref{KKRexpansion}), can be used to evaluate  
the connection as 
\begin{equation}
\begin{array}{ll}
{\bf \cal A}_n ({\bf k}) = i \int\limits_\omega \Psi_{n {\bf k}}^\dagger ({\bf r})
\boldsymbol\nabla_{\bf k} \Psi_{n {\bf k}} ({\bf r}) d {\bf r} + 
\int\limits_\omega \Psi_{n {\bf k}}^\dagger ({\bf r})
{\bf r} \Psi_{n {\bf k}} ({\bf r}) d {\bf r}\ .
\end{array}
\label{localconnection}
\end{equation}
From the point of view of the above discussion it should be stressed that the 
integrals in the above expression are over a chosen unit cell only and they 
are not the usual matrix elements between Bloch states. Clearly, such matrix 
elements would feature integrals over all the space with the corresponding orthogonality. In contrast, while integrating over a
unit cell the Bloch states are not orthogonal.

The purpose of writing  
${\bf \cal A}_{n}({\bf k})$ 
in the form of Eq.~(\ref{localconnection}) is not to attribute it to the Bloch states, but to facilitate 
its calculation using the local expansion of Bloch states given by Eq.~(\ref{KKRexpansion}).
As will become apparent shortly, the two contributions on r.h.s. of
Eq.~(\ref{localconnection}) correspond to different aspects of the problem. Therefore,
it is convenient to deal with them separately. For each reference we shall
call the first term ${\bf \cal A}_{n}^k ({\bf k})$
and the second ${\bf \cal A}_{n}^r ({\bf k})$.

Let us use the KKR expansion given by Eq.~(\ref{KKRexpansion}) and the fact that the 
scattering states $\Phi_Q ({\cal E}; {\bf r})$ can be normalized to 1 within a 
unit cell. Then, a straightforward calculation of ${\bf \cal A}_n^k ({\bf k})$ yields
\begin{equation}
\begin{array}{ll}
{\bf \cal A}_n^k ({\bf k}) = {\bf \cal A}_n^{KKR}({\bf k}) + {\bf \cal A}_n^v ({\bf k})\ ,
\end{array}
\label{kconnection}
\end{equation}
where (a detailed derivation is given in Appendix~\ref{App_b})
\begin{equation}
\begin{array}{ll}
{\bf \cal A}_n^v ({\bf k}) = i {\bf v}_n {\bar C}_n^\dagger \Bar{\Bar \Delta} {\bar C}_n =
- {\bf v}_n Im \{ {\bar C}_n^\dagger \Bar{\Bar \Delta} {\bar C}_n\}
\end{array}
\label{vconnection}
\end{equation}
with
\begin{equation}
\begin{array}{ll}
(\Bar{\Bar \Delta})_{Q Q^\prime} ({\cal E}) = \delta_{Q Q^\prime}
\int\limits_w  \Phi_Q^\dagger ({\cal E};{\bf r}) \frac{\partial \Phi_{Q^\prime} 
({\cal E};{\bf r})}{\partial {\cal E}} d {\bf r}
\end{array}
\label{Deltamatrix}
\end{equation}
and
\begin{equation}
\begin{array}{ll}
{\bf \cal A}_n^{KKR}({\bf k}) = i {\bar C}_n^\dagger  
\boldsymbol\nabla_{\bf k} {\bar C}_n = - Im \{ {\bar C}_n^\dagger  
\boldsymbol\nabla_{\bf k} {\bar C}_n\}\ . 
\end{array}
\label{KKRconnection}
\end{equation}
Here the matrix $\Bar{\Bar \Delta}$ is diagonal because the angular part of the 
KKR-basis set (Eq.~(\ref{basis_set})) does not depend on energy. Clearly, the term given 
by Eq.~(\ref{KKRconnection}) is similar to the standard formula for the connection. 
It is associated with the eigenvalue problem of Eq.~(\ref{eqno6}) in the usual way 
\cite{Berry84} and hence can be regarded as a property of the KKR matrix 
$\Bar{\Bar M} ({\cal E};{\bf k})$. This term is purely real since 
${\bar C}_n^\dagger \boldsymbol\nabla_{\bf k} {\bar C}_n$ is a purely imaginary quantity
due to the normalization ${\bar C}_n^\dagger {\bar C}_n =1$. The other term, 
${\bf \cal A}_n^v ({\bf k})$, is always parallel to the group velocity and purely real
due to the antihermitian property of the matrix $\Bar{\Bar \Delta}$.

Turning to the second term in Eq.~(\ref{localconnection}) and using the local expansion 
of Eq.~(\ref{KKRexpansion}) one readily finds
\begin{equation}
\begin{array}{ll}
{\bf \cal A}_n^r({\bf k}) = {\bar C}_n^\dagger ({\bf k}) \Bar{\Bar {{\bf r}}}  
{\bar C}_n ({\bf k})\ ,
\end{array}
\label{rconnection}
\end{equation}
where the vectorial matrix $\Bar{\Bar {\bf r}}$ is defined as
\begin{equation}
\begin{array}{ll}
({\Bar{\Bar {\bf r}}})_{Q Q^\prime} ({\cal E}) = \int\limits_\omega
\Phi_Q^\dagger ({\cal E};{\bf r}) {\bf r} \Phi_{Q^\prime} ({\cal E};{\bf r})
d {\bf r}\ . 
\end{array}
\label{rmatrix}
\end{equation}
Then the full connection is given by 
\begin{equation}
\begin{array}{ll}
{\bf \cal A}_n ({\bf k}) = {\bf \cal A}_n^{KKR}({\bf k}) + {\bf \cal A}_n^v ({\bf k}) +
{\bf \cal A}_n^r ({\bf k})
\end{array}
\label{fullconnection}
\end{equation}
together with Eqs.~(\ref{vconnection}),  (\ref{KKRconnection}), and (\ref{rconnection}).

The next subsection is devoted to present a method for calculating the 
curvature given by Eq.~(\ref{eqno(3)})  within this framework.

\subsection{KKR formula for Abelian Berry curvature}

It follows from Eq.~(\ref{fullconnection}) that the curvature can be considered 
as a sum of the following contributions
\begin{equation}
\begin{array}{ll}
\boldsymbol\Omega_n ({\bf k}) = \boldsymbol{\Omega }_{n}^{KKR} ({\bf k}) +
\boldsymbol\Omega_n^v ({\bf k}) + \boldsymbol\Omega_n^r ({\bf k})\ .  
\end{array}
\label{curvaturecorrection}
\end{equation}
We start with the first term of r.h.s. in the equation above,
namely $\boldsymbol\Omega_n^{KKR} ({\bf k}) = \boldsymbol\nabla_{\bf k} 
\times {\bf \cal A}_n^{KKR} ({\bf k})$. This is the curvature associated 
with the KKR eigenvalue problem of Eq.~(\ref{eqno6}). To deal with it, we note that 
\begin{equation}
\begin{array}{ll}
\boldsymbol\Omega_n^{KKR} ({\bf k}) = i \boldsymbol\nabla_{\bf k}
{\bar C}_n^\dagger \times \boldsymbol\nabla_{\bf k}{\bar C}_n =
- Im \{ \boldsymbol\nabla_{\bf k}
{\bar C}_n^\dagger \times \boldsymbol\nabla_{\bf k}{\bar C}_n\}\ .
\end{array}
\label{KKRcurvatureA}
\end{equation}
This is the standard form of the Berry curvature derived from a matrix eigenvalue problem.
\cite{Berry84}  However, because the KKR matrix $\Bar{\Bar M}
({\cal E};{\bf k})$ is not Hermitian, the algebra from here on deviates 
somewhat from the usual procedures.\cite{Berry84,Bohm03} In particular, the completeness relation 
$\sum_m {\bar C}_m {\bar C}_m^\dagger = \mathbf{\Bar{\Bar 1}}$ for $\Bar{\Bar M} ({\cal E};{\bf k})$ beeing Hermitian fails in our case. 
Instead, to transform 
Eq.~(\ref{KKRcurvatureA}) into a computationally convenient form, we must use 
\begin{equation}
\begin{array}{ll}
\sum\limits_{m=1}^{N} \frac{{\bar C}_m {\bar D}_m^\dagger}{{\bar D}_m^\dagger {\bar C}_m} 
= \sum\limits_{m=1}^{N} \frac{{\bar D}_m {\bar C}_m^\dagger}{{\bar C}_m^\dagger {\bar D}_m} 
= \mathbf{\Bar{\Bar 1}}\ ,
\end{array}
\label{completeness}
\end{equation}
where as before ${\bar C}_m$ and ${\bar D}_m$ are right and left eigenvectors of $\Bar{\Bar M}
({\cal E};{\bf k})$, respectively.\cite{Kalaba81}  Here the sum is going
over all $N$ eigenstates of the matrix $\Bar{\Bar M}$ that has a dimension of $N\times N$.  
Substituting Eq.~(\ref{completeness}) into Eq.~(\ref{KKRcurvatureA}) the KKR curvature takes 
the following form 
\begin{equation}
\begin{array}{ll}
\boldsymbol\Omega_n^{KKR} ({\bf k}) = - Im \{
\sum\limits_m \frac{\boldsymbol\nabla_{\bf k} {\bar C}_n^\dagger {\bar C}_m \times
{\bar D}_m^\dagger \boldsymbol\nabla_{\bf k} {\bar C}_n }{{\bar D}_m^\dagger {\bar C}_m}
\} \ .
\end{array}
\label{KKRcurvatureB}
\end{equation}

The next move is to eliminate the derivatives $\boldsymbol\nabla_{\bf k} {\bar C}_n$ in 
favor of $\boldsymbol\nabla_{\bf k} \Bar{\Bar M}$, similar to the case of the velocity formula 
in Eq.~(\ref{eqno8c}), by studying the gradient of Eq.~(\ref{eqno6}) with respect to
${\bf k}$. The details are given in Appendix \ref{App_c}. Here we merely record the result which 
facilitates the numerical evaluation $\boldsymbol\Omega_n^{KKR} ({\bf k})$ in 
Eq.~(\ref{KKRcurvatureB}):

\begin{equation}
\begin{array}{ll}
\boldsymbol\Omega_n^{KKR} ({\bf k}) = - Im \{
\sum\limits_{m \ne n} \frac 1{{\bar D}_m^\dagger {\bar C}_m
(\lambda_n - \lambda_m)} \\ \sum\limits_{k \ne n} 
\frac{[{\bar C}_k^\dagger {\bar C}_m - ({\bar C}_k^\dagger {\bar C}_n)
({\bar C}_n^\dagger {\bar C}_m) ]}{{\bar C}_k^\dagger {\bar D}_k} 
\frac{({\bar D}_k^\dagger \boldsymbol\nabla_{\bf k} \Bar{\Bar M} 
{\bar C}_n)^* \times {\bar D}_m^\dagger \boldsymbol\nabla_{\bf k} 
\Bar{\Bar M} {\bar C}_n} { (\lambda_n^* - \lambda_k^*)} \}\ ,
\end{array}
\label{KKRomega}
\end{equation}
where
\begin{equation}
\begin{array}{ll}
\boldsymbol\nabla_{\bf k} \Bar{\Bar M} = \frac{\partial { \Bar{\Bar G} ({\cal E};{\bf k})}}
{\partial {\bf k}} \Delta t({\cal E})\vert_{_{_{{\cal E}={\cal E}_n ({\bf k})}}} + \ \ \ \  
\\ \ \ \ \ \ \ \ \ \ + {\bf v}_n ({\bf k}) \left[ \frac{\partial {\Bar{\Bar G} 
({\cal E};{\bf k})}}{\partial {\cal E}} \Delta t({\cal E}) + {\Bar{\Bar G} ({\cal E};{\bf k})} 
\frac{\partial \Delta t ({\cal E})} {\partial {\cal E}} \right]\vert_{_{_{{\cal E} = 
{\cal E}_n ({\bf k})}}}
\end{array}
\label{deltaM}
\end{equation}
It is reassuring to note that for a Hermitian KKR matrix $\Bar{\Bar M}$, for which
${\bar D}_i = {\bar C}_i$ and ${\bar C}_i^\dagger {\bar C}_j = \delta_{ij}$, 
Eq.~(\ref{KKRomega}) reduces to its conventional form \cite{Berry84}
\begin{equation}
\begin{array}{ll}
\boldsymbol\Omega_n^{KKR} ({\bf k}) = -Im \{ \sum\limits_{m \ne n}
\frac{{\bar C}_n^\dagger \boldsymbol\nabla_{\bf k} \Bar{\Bar M} {\bar C}_m 
\times {\bar C}_m^\dag \boldsymbol\nabla_{\bf k} \Bar{\Bar M} {\bar C}_n}
{(\lambda_n -\lambda_m)^2}\}\ .
\end{array}
\end{equation}

From the point of view of the present paper, Eqs.~(\ref{KKRomega}) and (\ref{deltaM}) 
together are one of our two central formal results. It expresses the contribution 
$\boldsymbol\Omega_n^{KKR} ({\bf k})$ to the Berry curvature in terms of the left and 
right eigenvectors, the group velocity, and  ${\bf k}$ and ${\cal E}$ derivatives of 
the KKR matrix. As will be demonstrated, these relations provide an efficient 
way of calculating $\boldsymbol\Omega_n^{KKR} ({\bf k})$ similarly to the manner of 
Eq.~(\ref{eqno8c}) done for the group velocity. The main difference is that now we need
the total derivative $\boldsymbol\nabla_{\bf k} \Bar{\Bar M}$ in Eq.~(\ref{deltaM})
instead of the partial derivative $\partial \Bar{\Bar M} / \partial {\bf k}$ used in 
Eq.~(\ref{eqno8c}).

Let us consider the term $\boldsymbol\Omega_n^v ({\bf k}) = \boldsymbol\nabla_{\bf k} 
\times {\bf \cal A}_n^v ({\bf k})$. A detailed derivation performed in Appendix \ref{App_c}, finally gives 
the following expression  
\begin{equation}
\begin{array}{ll}
\boldsymbol\Omega_n^v ({\bf k}) = 2 {\bf v}_n ({\bf k})\times \\
\ \ \ 
Im \{\sum\limits_{m \ne n} \frac{ [ {\bar C}_n^\dagger \Bar{\Bar \Delta} {\bar C}_m -
{\bar C}_n^\dagger \Bar{\Bar \Delta} {\bar C}_n ({\bar C}_n^\dagger {\bar C}_m) ]
{\bar D}_m^\dagger \boldsymbol\nabla_{\bf k} \Bar{\Bar M} {\bar C}_n}
{{\bar D}_m^\dagger {\bar C}_m (\lambda_n - \lambda_m)} \}\ ,
\end{array}
\label{Omegav}
\end{equation}
where the matrix $\Bar{\Bar \Delta}$ is defined by Eq.~(\ref{Deltamatrix}).
Due to the cross vector product, this term does not contribute to the Fermi surface integrals 
needed in calculations which follow Haldane's proposal.\cite{Haldane04}

Returning to the contribution $\boldsymbol\Omega_n^r ({\bf k})$
defined in Eq.~(\ref{curvaturecorrection}), on taking the curl, one finds
\begin{equation}
\begin{array}{ll}
\boldsymbol\Omega_n^r ({\bf k}) = \int\limits_\omega
\left[ ( \boldsymbol\nabla_{\bf k} | \Psi_{n {\bf k}} ({\bf r}) | ^{2} )
\times {\bf r} \right] d {\bf r} = \\ \ \ \ \ \ \ \ \ \ \ \ \ \ \ \ \ \ \ 
= 2 Re \{ \int\limits_\omega [ \Psi_{n {\bf k}}^\dagger ({\bf r})  
(\boldsymbol\nabla_{\bf k} \Psi_{n {\bf k}} ({\bf r})) \times {\bf r} ] 
d {\bf r} \}\ .
\end{array}
\label{rcurvature}
\end{equation}
Then, using the KKR expansion of Eq.~(\ref{KKRexpansion}) it follows (Appendix \ref{App_c}) that
\begin{equation}
\begin{array}{ll}
\boldsymbol\Omega_n^r ({\bf k}) = 2\cdot {\bf v}_n ({\bf k}) \times Re 
\{ {\bar C}_n^\dagger {\Bar {\Bar {\bf r}}}_{\cal E} {\bar C}_n \} - \\ -
2\cdot Re \{ \sum\limits_{m \ne n} \frac{ [ {\bar C}_n^\dagger {\Bar {\Bar {\bf r}}} 
{\bar C}_m - {\bar C}_n^\dagger {\Bar{\Bar{\bf r}}} {\bar C}_n ({\bar C}_n^\dagger {\bar C}_m) ] 
\times {\bar D}_m^\dagger \boldsymbol\nabla_{\bf k} \Bar{\Bar M} {\bar C}_n}
{{\bar D}_m^\dagger {\bar C}_m (\lambda_n - \lambda_m)} \}\ .
\end{array}
\label{rmatcurvature}
\end{equation}
Here the vectorial matrix ${\Bar {\Bar {\bf r}}}$ is given by Eq.~(\ref{rmatrix}),
and the vectorial matrix ${\Bar {\Bar {\bf r}}}_{\cal E}$ is defined as 
\begin{equation}
\begin{array}{ll}
({\Bar {\Bar {\bf r}}}_{\cal E})_{Q Q^\prime} ({\cal E}) = \int\limits_\omega
\Phi_Q^\dagger ({\cal E};{\bf r}) {\bf r} \frac{\partial \Phi_{Q^\prime} ({\cal E};{\bf r})}
{\partial {\cal E}} d {\bf r}\ . 
\end{array}
\label{rematrix}
\end{equation}

To summarize the above discussion, we note that the formula for the full curvature 
(Eq.~(\ref{curvaturecorrection})) together with Eqs.~(\ref{KKRomega}), (\ref{Omegav}), 
and (\ref{rmatcurvature}) constitutes a basis for calculating the conventionally defined 
$\boldsymbol\Omega_n ({\bf k})$. Here we used the eigenvectors and eigenvalues of the KKR matrix and the 
matrix elements with respect to the local scattering states $\Phi_Q ({\cal E}; {\bf r})$ given by Eqs.~(\ref{Deltamatrix}), (\ref{rmatrix}), and (\ref{rematrix}).
Clearly, these quantities are readily available in a KKR calculation which is aimed 
at computing the wavefunctions as well as the dispersion relation.\cite{Gradhand09}

An important point to mention is that the projection of $\boldsymbol\Omega_n ({\bf k})$
along the group velocity
\begin{equation}
\begin{array}{ll}
\boldsymbol\Omega_n ({\bf k}) \cdot {\bf v}_n ({\bf k}) = - {\bf v}_n ({\bf k}) \cdot
Im \{ \sum\limits_{m \ne n} \frac 1{{\bar D}_m^\dagger {\bar C}_m
(\lambda_n - \lambda_m)} \\ \sum\limits_{k \ne n} 
\frac{[{\bar C}_k^\dagger {\bar C}_m - ({\bar C}_k^\dagger {\bar C}_n)
({\bar C}_n^\dagger {\bar C}_m) ]}{{\bar C}_k^\dagger {\bar D}_k} 
\frac{({\bar D}_k^\dagger (\partial \Bar{\Bar M} / \partial {\bf k}) 
{\bar C}_n)^* \times {\bar D}_m^\dagger (\partial \Bar{\Bar M} / \partial {\bf k}) 
{\bar C}_n}{ (\lambda_n^* - \lambda_k^*)} \} \\ - 2 {\bf v}_n ({\bf k}) \cdot 
Re \{ \sum\limits_{m \ne n} \frac{ [ {\bar C}_n^\dagger {\Bar {\Bar {\bf r}}} {\bar C}_m -
{\bar C}_n^\dagger {\Bar {\Bar {\bf r}}} {\bar C}_n ({\bar C}_n^\dagger {\bar C}_m) ]
\times {\bar D}_m^\dagger (\partial \Bar{\Bar M} / \partial {\bf k}) {\bar C}_n}
{{\bar D}_m^\dagger {\bar C}_m (\lambda_n - \lambda_m)} \}
\end{array}
\label{projection}
\end{equation}
does not have any terms connected with the energy derivative.
This is then the second principle result of the current section. Its significance is
that no Fermi surface integral \emph{contains energy derivatives}.
Therefore, for calculating the anomalous Hall conductivity according to Haldane's approach~\cite{Haldane04},
one does not need a numerical differentiation at all, since the partial derivative $\partial \Bar{\Bar M} / \partial {\bf k}$ 
has to be taken analytically according to 
Eq.~(\ref{eqno8d}). 

Up to now we discussed the conventional Abelian case when there is no degeneracy
of the electronic states. In the next section we consider the Berry curvature in a general 
non-Abelian case.

\subsection{KKR formula for non-Abelian Berry curvature}

As it was discussed in detail by Shindou and Imura (Ref.~\onlinecite{Shindou05}),
the presence of degenerate Bloch bands makes the Berry curvature \emph{non-Abelian}. Since 
two covariant derivatives (with respect to ${\bf k}$) along different axes do not commute with
each other in the subspace spanned by the degenerate bands, the Abelian description fails. Namely, in the case of an $L$-fold
degeneracy the Berry curvature is not any more a vector, but a vector-valued matrix in 
$L$-dimensional space labeled $\Sigma$. Those elements can be written as \cite{Shindou05,Falko} 
\begin{equation}
\begin{array}{ll}
\boldsymbol\Omega_{ij} ({\bf k}) = i \langle \boldsymbol\nabla_{\bf k} u_{i {\bf k}} | \times |
\boldsymbol\nabla_{\bf k} u_{j {\bf k}} \rangle - \\ - i \sum\limits_{l \in \Sigma} \langle 
\boldsymbol\nabla_{\bf k} u_{i {\bf k}} |  u_{l {\bf k}} \rangle \times \langle u_{l {\bf k}} | 
\boldsymbol\nabla_{\bf k} u_{j {\bf k}} \rangle\ ,
\end{array}
\label{Omega_non_A}
\end{equation}
where indices $i$ and $j$ mean any two states from the set $\Sigma = \{1, 2, ..., L \}$ of 
the degenerate states. Below we derive detailed expressions for the non-Abelian Berry curvature 
given by Eq.~(\ref{Omega_non_A}) within the KKR method.

For typographical simplicity, in this subsection we use the inner products which will always 
mean an integration over the unit-cell. Then, using the Bloch theorem, 
 Eq.~(\ref{Omega_non_A}) can be rewritten as (omitting index ${\bf k}$ for the wave functions)  
\begin{widetext}
\begin{equation}
\begin{array}{ll}
\boldsymbol\Omega_{ij} ({\bf k}) = i \langle \boldsymbol\nabla_{\bf k} \Psi_i | \times |  
\boldsymbol\nabla_{\bf k} \Psi_j \rangle + \langle \boldsymbol\nabla_{\bf k} \Psi_i \times 
{\bf r} | \Psi_j \rangle - \langle \Psi_i | {\bf r} \times 
\boldsymbol\nabla_{\bf k} \Psi_j \rangle - \sum\limits_{l \in \Sigma}
\left\{ i \langle \boldsymbol\nabla_{\bf k} \Psi_i |  \Psi_l \rangle \times
\langle \Psi_l | \boldsymbol\nabla_{\bf k} \Psi_j \rangle - \right. \\ \left. 
\ \ \ \ \ \ \ \ \ \ \ \ \ \ \ \ \ \ \ \ \ 
- \langle \Psi_i |{\bf r}| \Psi_l \rangle \times 
\langle \Psi_l | \boldsymbol\nabla_{\bf k} \Psi_j \rangle +
\langle \boldsymbol\nabla_{\bf k} \Psi_i |  \Psi_l \rangle \times
\langle \Psi_l |{\bf r}| \Psi_j \rangle + 
i \langle \Psi_i |{\bf r}| \Psi_l \rangle \times \langle \Psi_l |{\bf r}| \Psi_j \rangle
\right\}\ .
\end{array}
\end{equation}
Similar to the previous subsection, we can generalize our separation of the Berry curvature into the following contributions 
\begin{equation}
\begin{array}{rr}
\boldsymbol\Omega_{ij} ({\bf k}) = \boldsymbol\Omega_{ij}^k ({\bf k}) +
\boldsymbol\Omega_{ij}^r ({\bf k}) = \boldsymbol\Omega_{ij}^{KKR} ({\bf k}) +
\boldsymbol\Omega_{ij}^v ({\bf k}) + \boldsymbol\Omega_{ij}^r ({\bf k})\ .
\end{array}
\label{Omega_ij}
\end{equation}
Here 
$\boldsymbol\Omega_{ij}^k ({\bf k})$ splits into 
\begin{equation}
\begin{array}{ll}
\boldsymbol\Omega_{ij}^{KKR} ({\bf k}) = i \sum\limits_{m \notin \Sigma}
\sum\limits_{k \notin \Sigma} \frac{[{\bar C}_k^\dagger {\bar C}_m -
\sum_{l \in \Sigma} ({\bar C}_k^\dagger {\bar C}_l)
({\bar C}_l^\dagger {\bar C}_m)] ({\bar D}_k^\dagger \boldsymbol\nabla_{\bf k} \Bar{\Bar M}
{\bar C}_i)^* \times {\bar D}_m^\dagger \boldsymbol\nabla_{\bf k} \Bar{\Bar M} {\bar C}_j}
{{\bar C}_k^\dagger {\bar D}_k (\lambda_i^* - \lambda_k^*)
{\bar D}_m^\dagger {\bar C}_m (\lambda_j - \lambda_m)}\ 
\end{array}
\label{Curv_KKR_nonA}
\end{equation}
and
\begin{equation}
\begin{array}{rrl}
\boldsymbol\Omega_{ij}^v ({\bf k}) = -i &\{& {\bf v}_i \times \sum\limits_{m \notin \Sigma}
\frac{[{\bar C}_i^\dagger \Bar{\Bar \Delta} {\bar C}_m - \sum_{l \in \Sigma}
({\bar C}_i^\dagger \Bar{\Bar \Delta} {\bar C}_l)({\bar C}_l^\dagger {\bar C}_m) ]
{\bar D}_m^\dagger \boldsymbol\nabla_{\bf k} \Bar{\Bar M}
{\bar C}_j}{{\bar D}_m^\dagger {\bar C}_m (\lambda_j - \lambda_m)}\  +\\
&+&{\bf v}_j \times \sum\limits_{m \notin \Sigma}
\frac{[{\bar C}_m^\dagger \Bar{\Bar \Delta} {\bar C}_j - \sum_{l \in \Sigma}
({\bar C}_m^\dagger {\bar C}_l) ({\bar C}_l^\dagger \Bar{\Bar \Delta} {\bar C}_j) ]
({\bar D}_m^\dagger \boldsymbol\nabla_{\bf k}
\Bar{\Bar M} {\bar C}_i)^*}{{\bar C}_m^\dagger {\bar D}_m (\lambda_i^* - \lambda_m^*)}\} \ + \\
&+&i\left[{\bf v}_i\times {\bf v}_j\right]\left\{\bar{c}_i^{\dagger}\bar{\bar \Delta}_{\cal E} \bar{c}_j-
\sum\limits_{l\in \Sigma}(\bar{c}_i^{\dagger}\bar{\bar \Delta}^{\dagger} \bar{c}_l)(\bar{c}_l^{\dagger}\bar{\bar \Delta} \bar{c}_j)\right\}
\end{array}
\label{Curv_v_nonA}
\end{equation}
with the new matrix 
$\Bar{\Bar \Delta}_{\cal E}$ defined as
\begin{equation}
\begin{array}{ll}
(\Bar{\Bar \Delta}_{\cal E})_{Q Q^\prime} ({\cal E}) = \int\limits_w  \frac{\partial\Phi_Q^\dagger ({\cal E};{\bf r})}{\partial {\cal E}} \frac{\partial \Phi_{Q^\prime} 
({\cal E};{\bf r})}{\partial {\cal E}} d {\bf r}\ .
\end{array}
\label{DeltaEmatrix}
\end{equation}
The last term in Eq.~(\ref{Omega_ij}) can be written as 
\begin{equation}
\begin{array}{rrl}
\boldsymbol\Omega_{ij}^r ({\bf k}) = &{\bf v}_i& \times [ {\bar C}_i^\dagger
{\Bar {\Bar {\bf r}}}_{\cal E}^\dagger {\bar C}_j - \sum\limits_{l \in \Sigma}
({\bar C}_i^\dagger \Bar{\Bar \Delta}^\dagger {\bar C}_l) ({\bar C}_l^\dagger
{\Bar{\Bar {\bf r}}} {\bar C}_j) ]\ +\\
+ &{\bf v}_j& \times [ {\bar C}_i^\dagger
{\Bar {\Bar {\bf r}}}_{\cal E} {\bar C}_j -  \sum\limits_{l \in \Sigma}
({\bar C}_i^\dagger {\Bar{\Bar {\bf r}}} {\bar C}_l)
({\bar C}_l^\dagger \Bar{\Bar \Delta} {\bar C}_j) ]  
- i \sum\limits_{l \in \Sigma} ({\bar C}_i^\dagger {\Bar{\Bar {\bf r}}} {\bar C}_l) \times
({\bar C}_l^\dagger {\Bar{\Bar {\bf r}}} {\bar C}_j)\ - \\ 
-&\sum\limits_{m \notin \Sigma}& \frac{[{\bar C}_m^\dagger {\Bar{\Bar {\bf r}}} {\bar C}_j -
\sum_{l \in \Sigma}
({\bar C}_m^\dagger {\bar C}_l) {\bar C}_l^\dagger {\Bar{\Bar {\bf r}}} {\bar C}_j ]
\times({\bar D}_m^\dagger \boldsymbol\nabla_{\bf k} \Bar{\Bar M}
{\bar C}_i)^*}{{\bar C}_m^\dagger {\bar D}_m (\lambda_i^* - \lambda_m^*)}\ -\\
-&\sum\limits_{m \notin \Sigma}& \frac{[{\bar C}_i^\dagger {\Bar{\Bar {\bf r}}} {\bar C}_m -
\sum_{l \in \Sigma}
{\bar C}_i^\dagger {\Bar{\Bar {\bf r}}} {\bar C}_l ({\bar C}_l^\dagger {\bar C}_m)]
\times {\bar D}_m^\dagger \boldsymbol\nabla_{\bf k} \Bar{\Bar M}
{\bar C}_j}{{\bar D}_m^\dagger {\bar C}_m (\lambda_j - \lambda_m)}\ . \ \ \ \ 
\end{array}
\label{Curv_r_nonA}
\end{equation}
\end{widetext}
A detailed derivation of these formulas is given in Appendix \ref{App_d}. To get the expressions for
the Abelian case, obtained in the previous section, one needs to consider the diagonal element
$\boldsymbol\Omega_{ii} ({\bf k})$ and restrict the sum over $l$ just to the term $l=i$.

\section{Results for the Berry curvature}

Here we present the results for the Berry curvature at the Fermi surface 
of Al, Cu, Au, and Pt bulk crystals. All of them are non-magnetic materials with space 
inversion symmetry. As was mentioned in Section I, in such a case the electron states are 
two-fold degenerate at each {\bf k} point. In other words, they form a Kramers doublet. Therefore, according to the discussion of 
Section III, the non-Abelian Berry curvature is a vector-valued matrix in the two-dimensional space of the two degenerate bands.

In general, each matrix element of 
$\boldsymbol\Omega ({\bf k})$ is gauge dependent. It would be meaningless to visualize the elements for an arbitrary gauge. 
A gauge-independent quantity is the vector
$Tr\left[\boldsymbol\Omega ({\bf k})\right]$, but it vanishes for Kramers-degenerate bands. 
Other gauge independent quantities are
$Tr\left[S^\mu({\bf k})\Omega^\mu({\bf k})\right]$ with $\mu=x,y,z$ 
using the spin matrices
$S_{ij}^\mu({\bf k})=\left<\Psi_i|\beta\sigma_\mu|\Psi_j\right>$ in the subspace spanned by the two degenerate bands. 
However, these quantities combine already two effects stemming from the Berry curvature as well as the spin mixing of the wave functions.~\cite{Gradhand09} 
Hence, some features of the Berry curvature may be hidden.

Since there is no convenient gauge invariant quantity to plot we have chosen a physically appealing gauge. Similarly to what was discussed in Ref.~\onlinecite{Gradhand09}, 
it is a special linear combination of the degenerate states such that the off-diagonal matrix 
elements of the spin operator $\Sigma_z=\beta\sigma_z$  in the subspace of the two degenerate 
bands are zero. Such a transformation can always be performed. For simplicity we present the Berry 
curvature for one of the degenerate bands only, since for the diagonal elements the relation
$\boldsymbol\Omega_{11}=-\boldsymbol\Omega_{22}$ holds. 
The off-diagonal terms are more complicated being not purely real, but complex numbers.

\begin{figure}[ht]
\includegraphics[width=\LL]{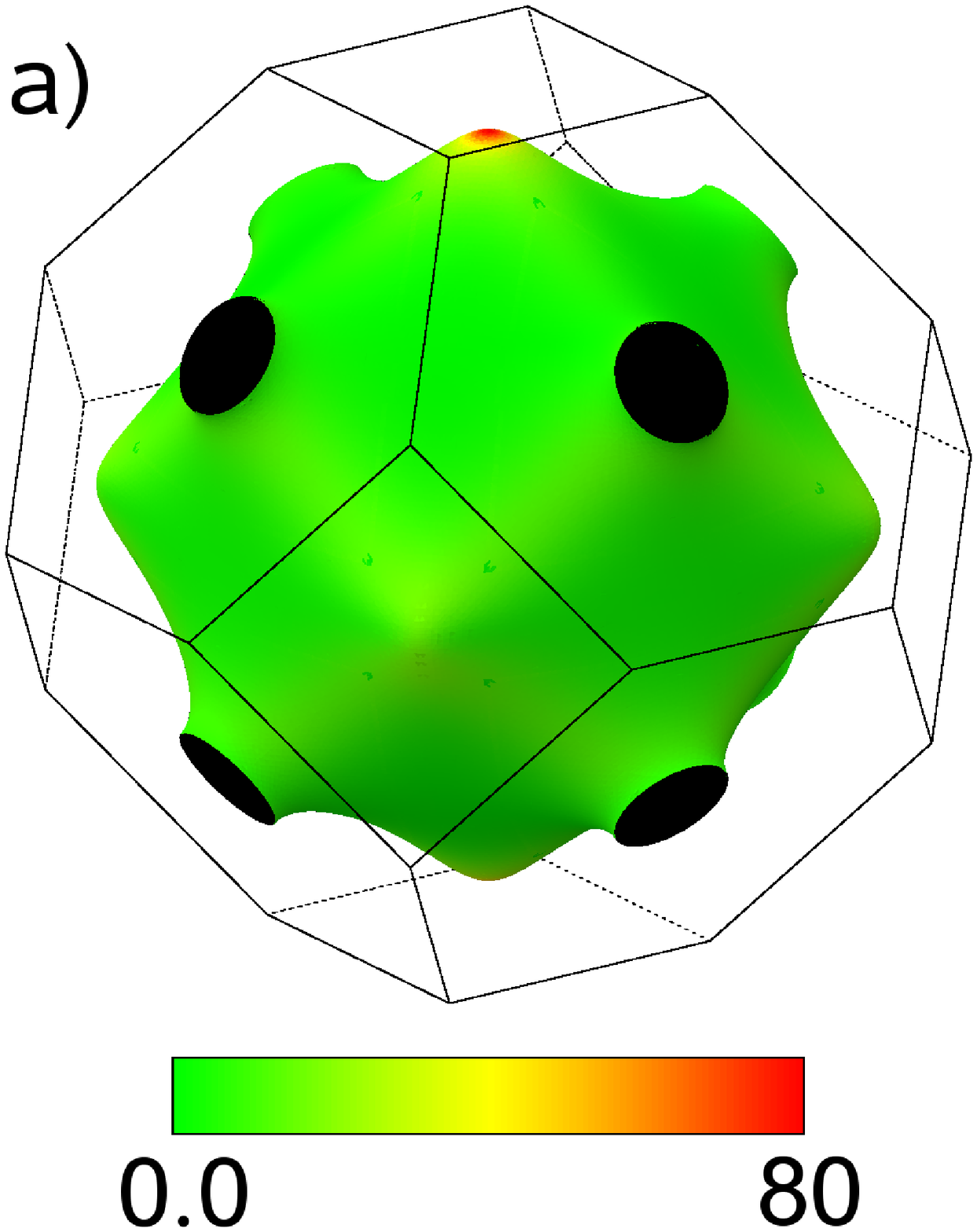}
\includegraphics[width=\LL]{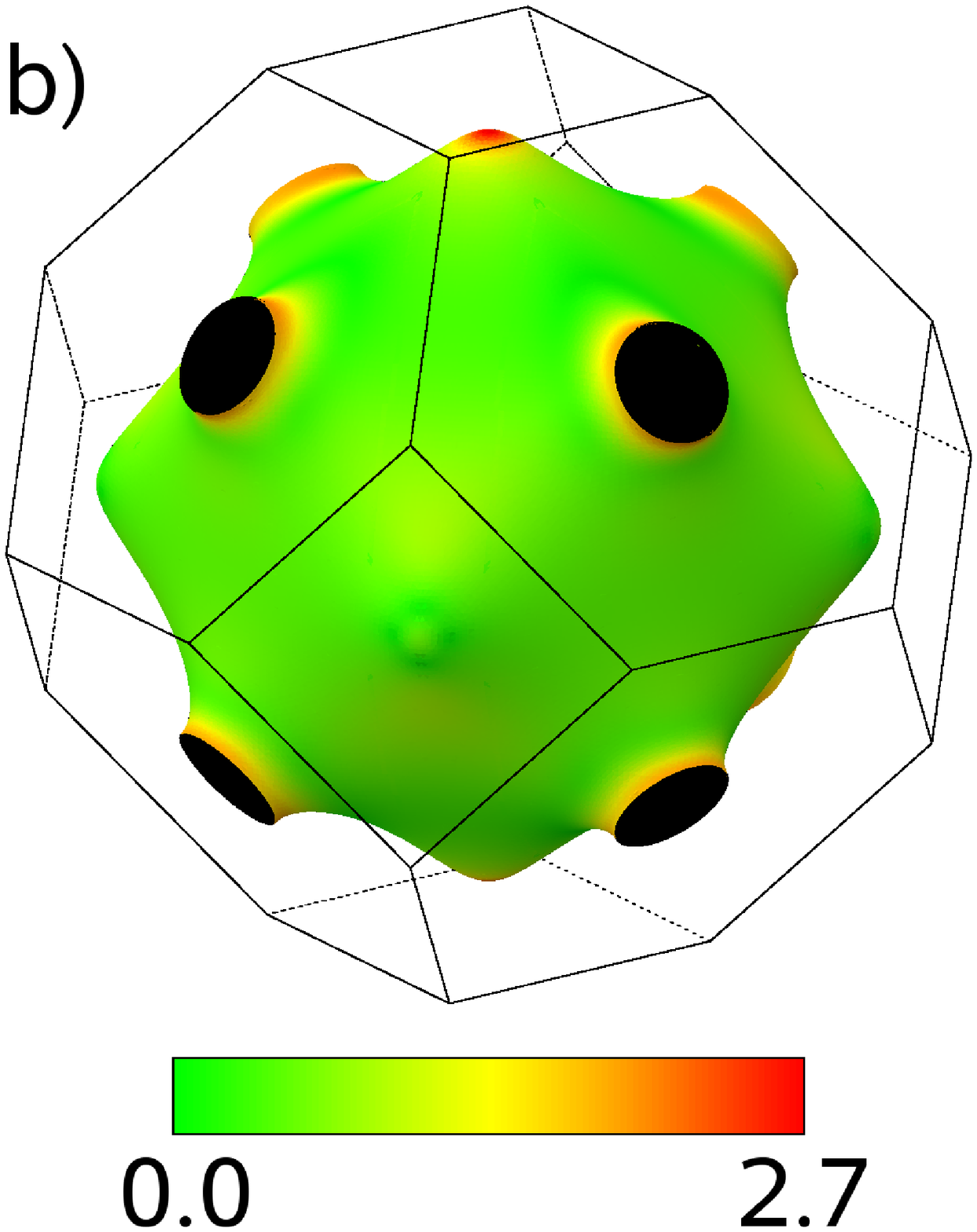}
\includegraphics[width=\LL]{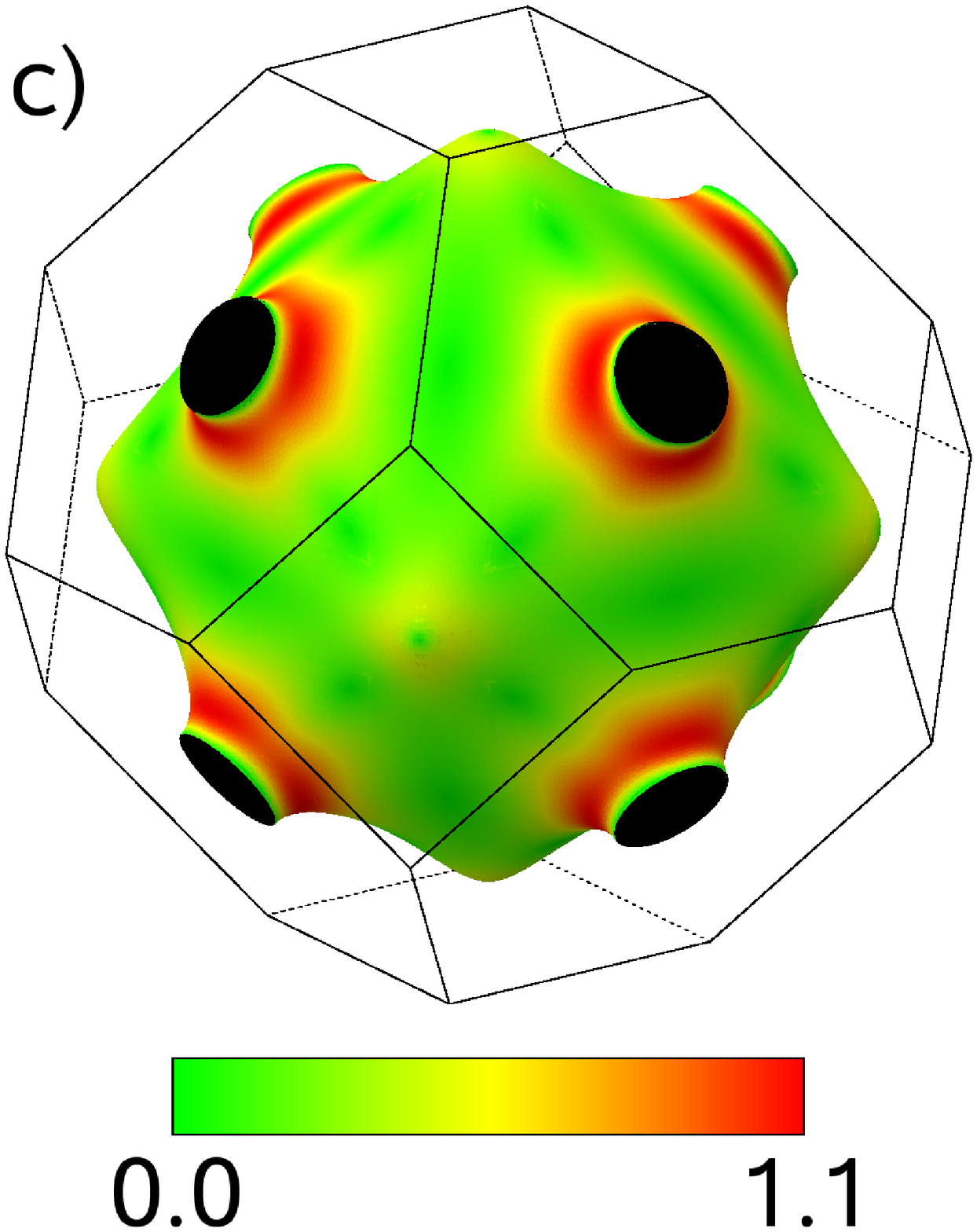}
\caption{The length of the diagonal components of the Berry curvature on the Fermi surface of Au  in a.u.: 
a) the KKR part $\boldsymbol{\Omega}_{ii}^{KKR}({\bf k})$ according to Eq.~(\ref{Curv_KKR_nonA}), 
b) the contribution from $\boldsymbol{\Omega}_{ii}^r({\bf k})$ given by 
Eq.~(\ref{Curv_r_nonA}), and c) the part $\boldsymbol{\Omega}_{ii}^v({\bf k})$ introduced by the energy dependence of the basis functions 
according to Eq.~(\ref{Curv_v_nonA}).}
\label{fig.:berry_Au_all}
\end{figure}

\begin{figure*}[ht]
\includegraphics[width=\LL]{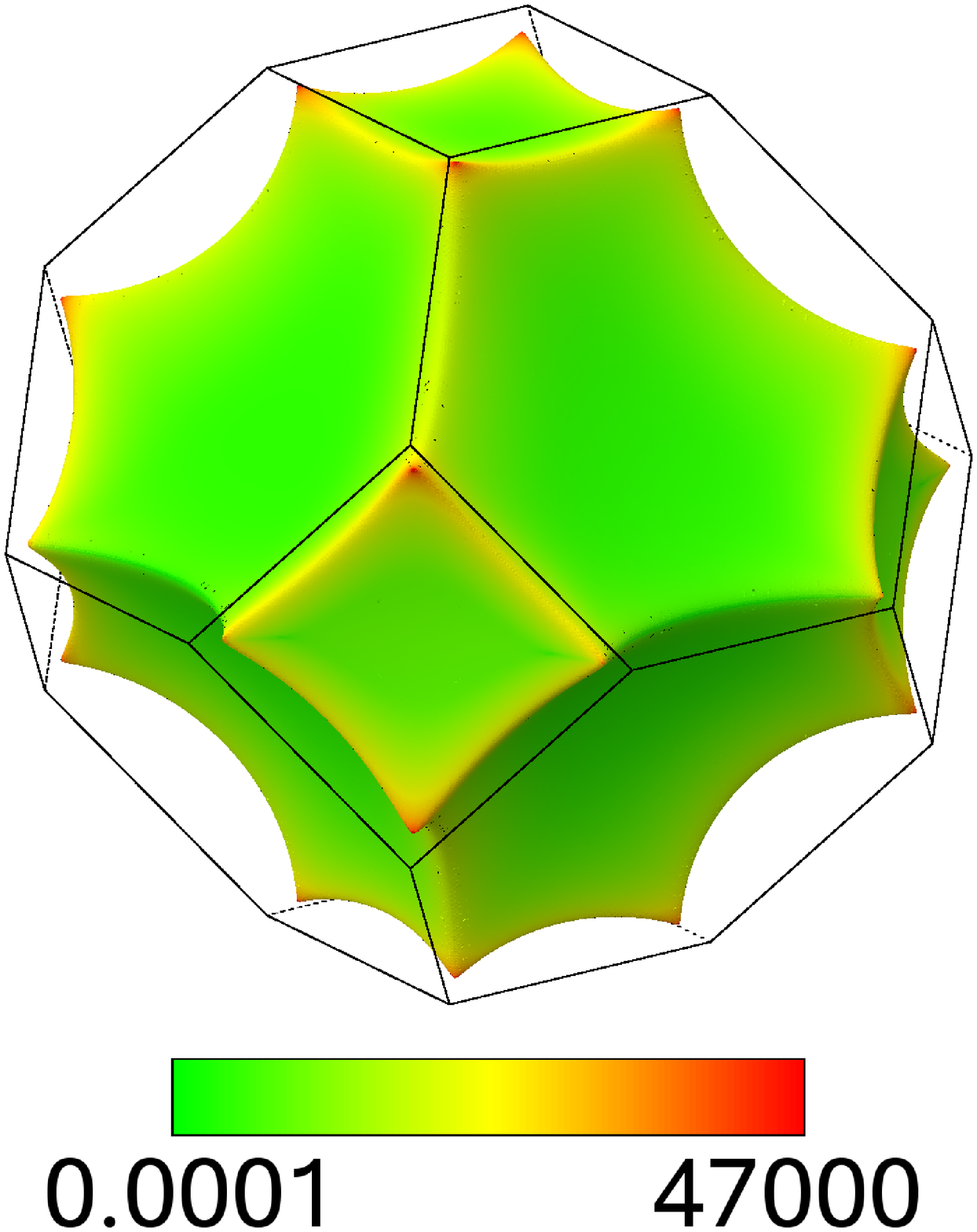}
\includegraphics[width=\LL]{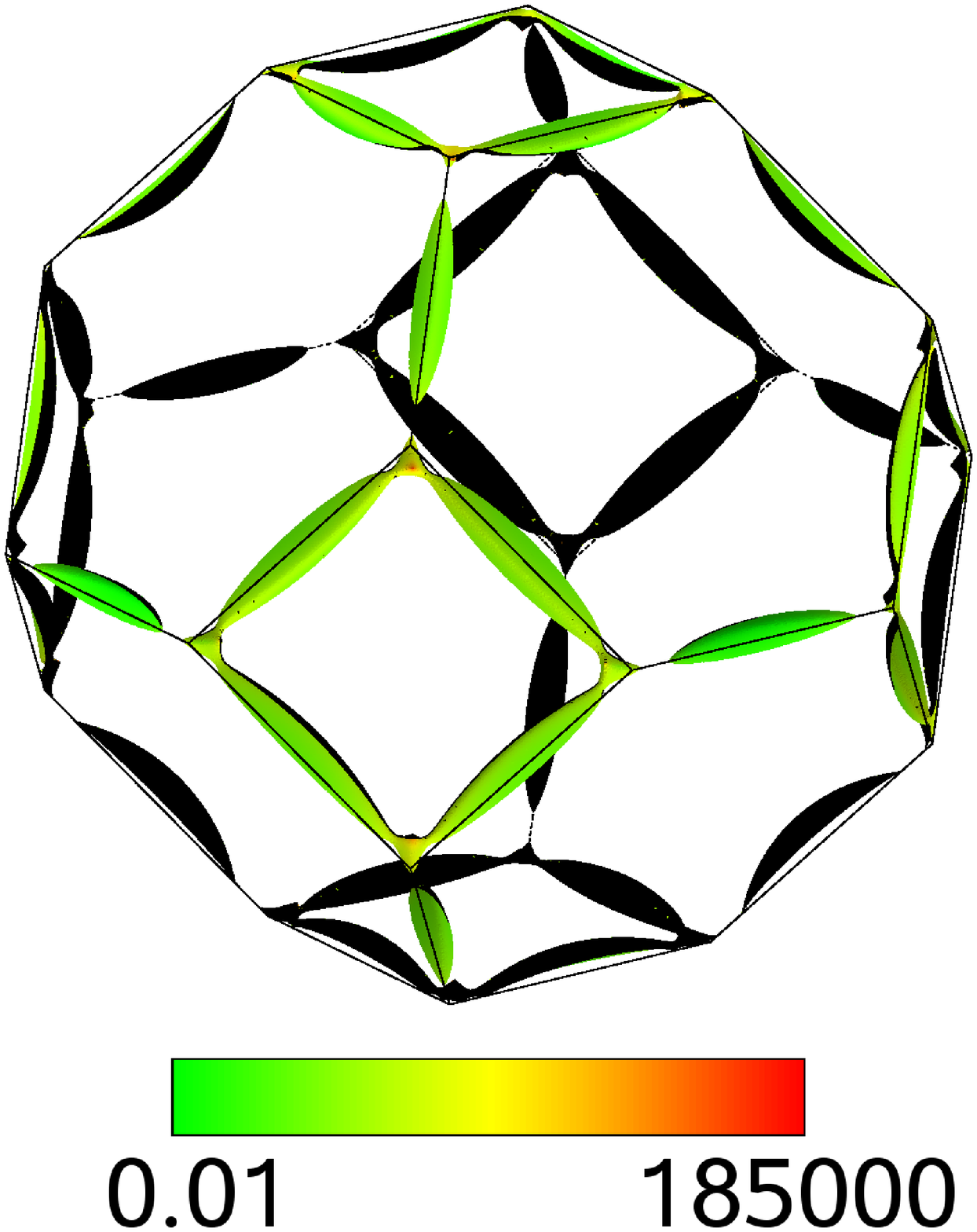}
\includegraphics[width=\LL]{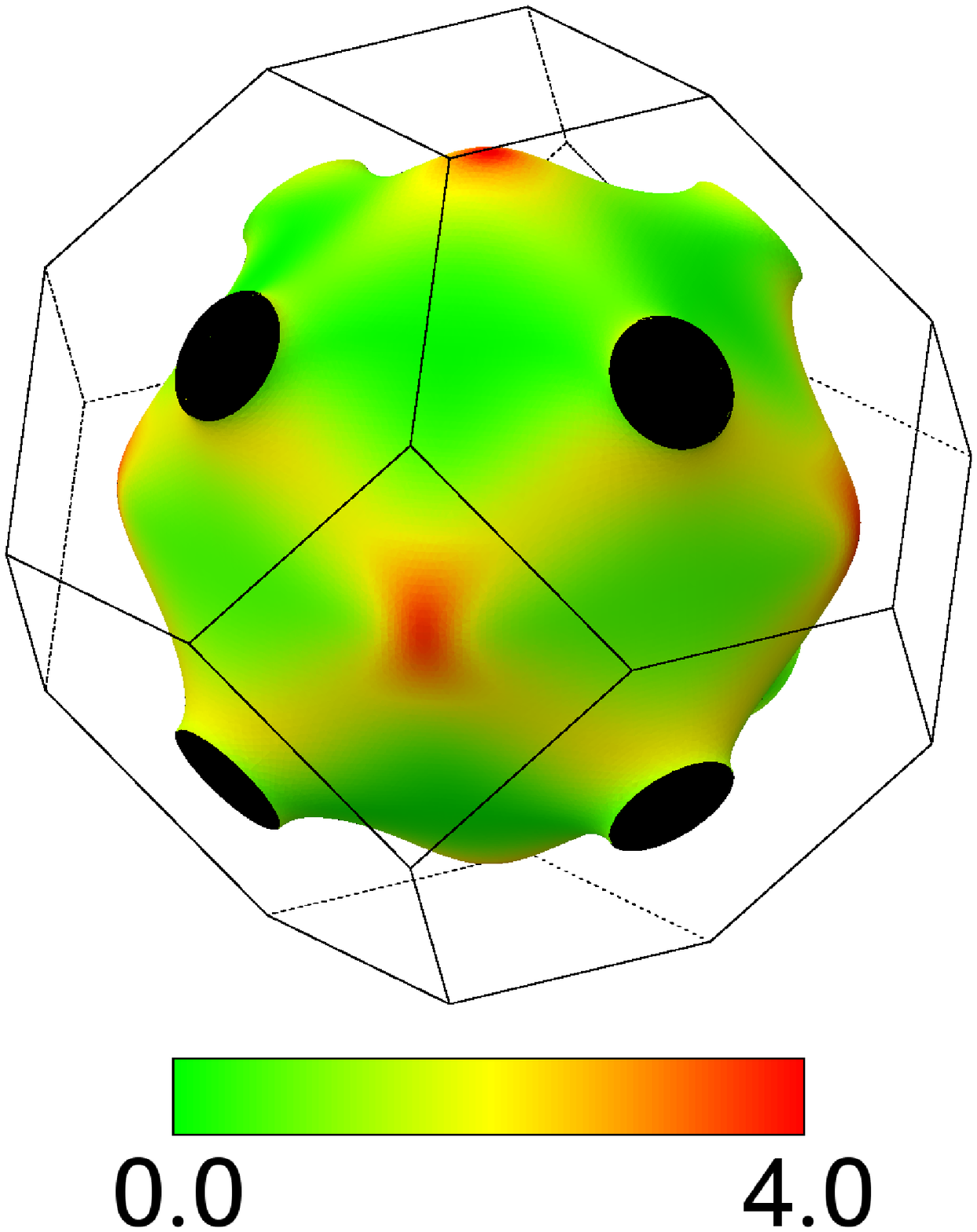}
\includegraphics[width=\LL]{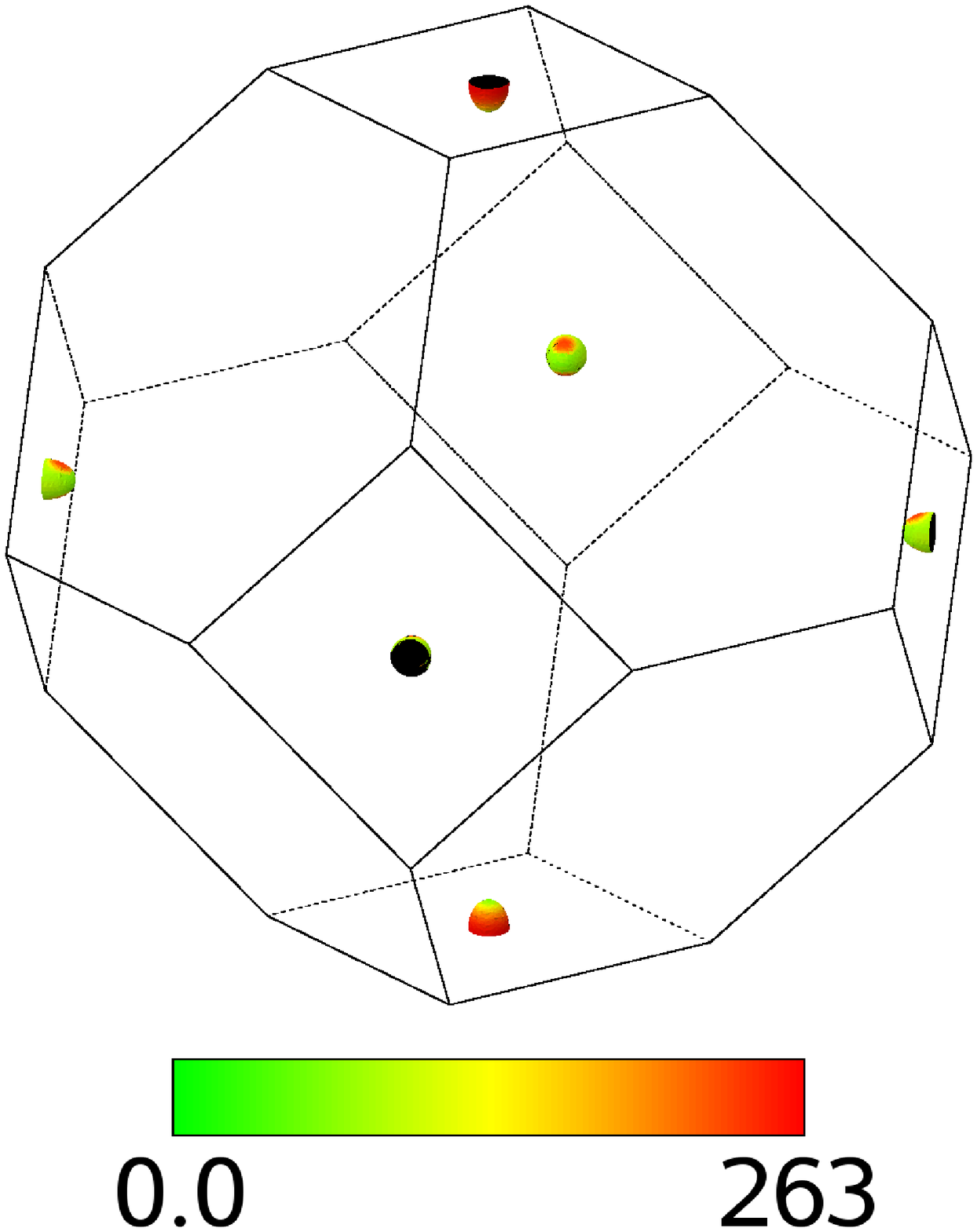}
\includegraphics[width=\LL]{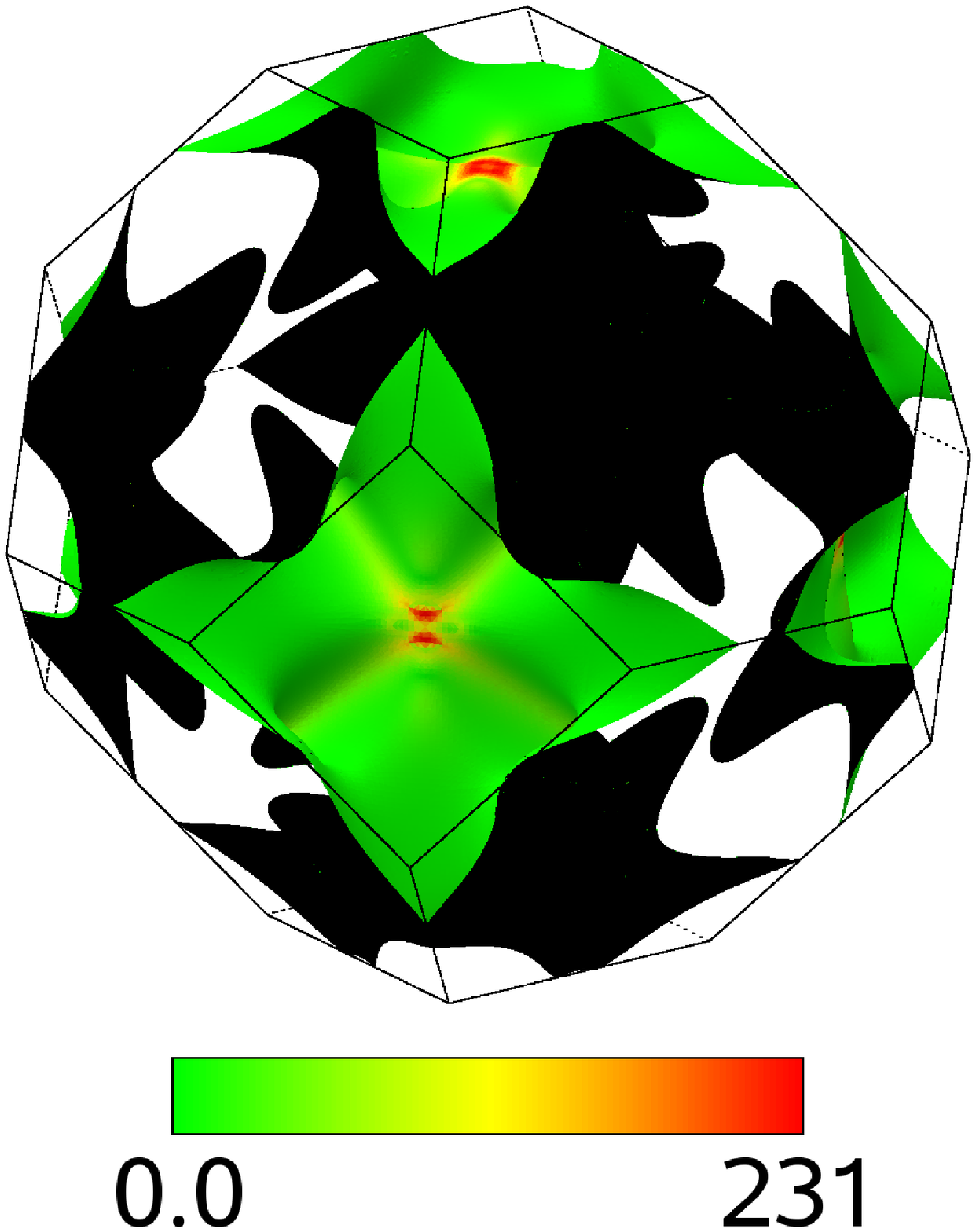}
\includegraphics[width=\LL]{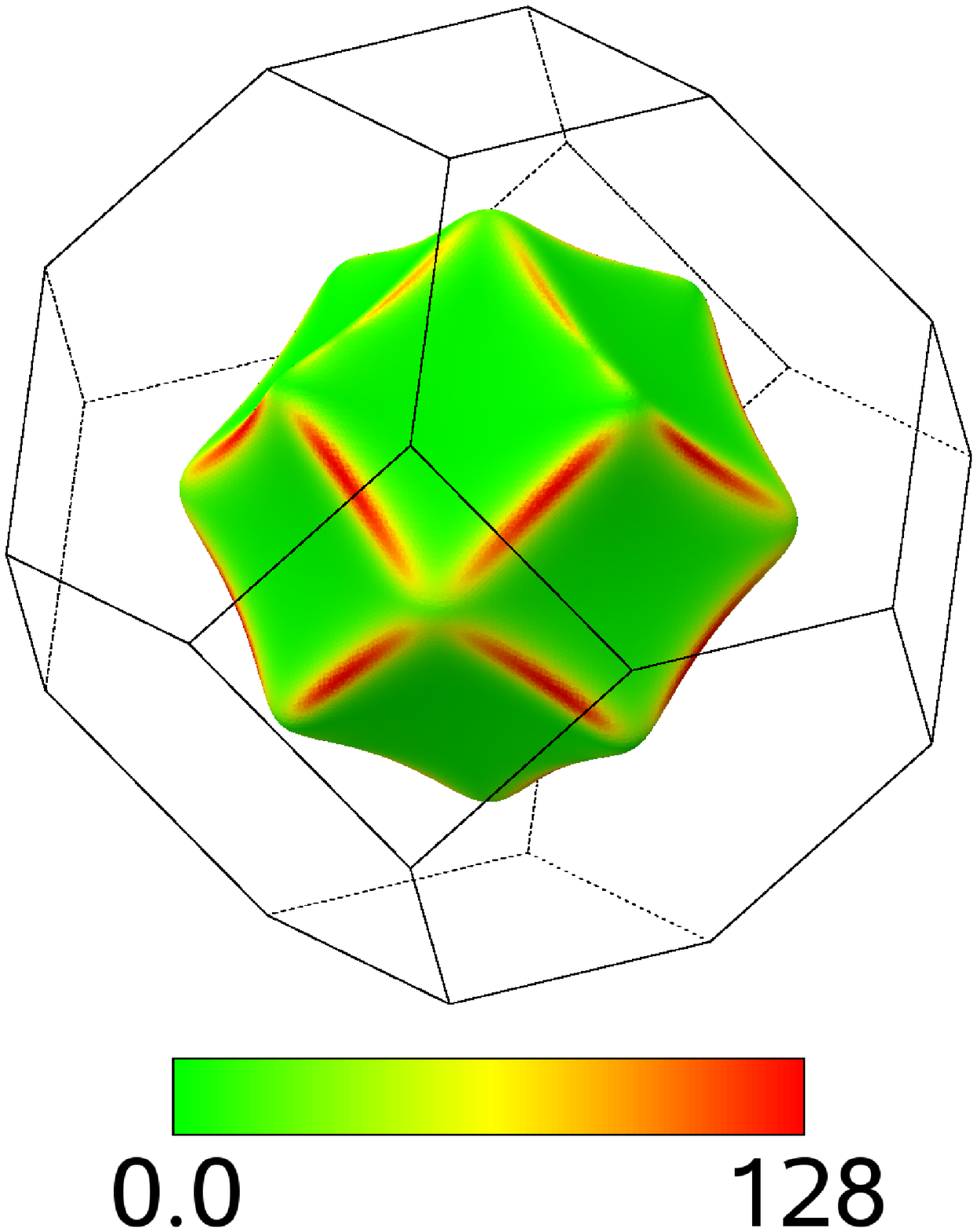}
\caption{The absolute value (in a.u.) of the diagonal component of the Berry curvature for the Fermi surface of several metals. 
From left to right: Al (3rd and 5th band), Cu (11th band), and Pt (7th, 9th, and 11th band). 
For Al we used a logarithmic scale to visualize the important regions.}\label{fig.:berry_all}
\end{figure*}

In Fig.~\ref{fig.:berry_Au_all} we compare the three separate parts 
contributing to the Berry curvature from Eqs.~(\ref{Curv_KKR_nonA}), (\ref{Curv_r_nonA}), and (\ref{Curv_v_nonA}).
The first one (Fig.~\ref{fig.:berry_Au_all}~(a)) is the KKR part 
$\boldsymbol{\Omega}^{KKR}({\bf k})$ that clearly has the dominant contribution. 
The maximum value of the contribution $\boldsymbol{\Omega}^{r}({\bf k})$ in Fig.~\ref{fig.:berry_Au_all}~(b) is less than $4\%$ of 
$\boldsymbol{\Omega}^{KKR}({\bf k})$, and $\boldsymbol{\Omega}^{v}({\bf k})$ shown in Fig.~\ref{fig.:berry_Au_all}~(c) contributes less than 2\%.
The same holds for all the other considered systems for which only the total Berry curvature $\Omega({\bf k})$ is summarized in Fig.~\ref{fig.:berry_all}.

Here, the interesting result is that the maximum value for the length of the Berry curvature is largest for Al which is actually the lightest 
element with the weakest atomic spin-orbit coupling. 
However, the region of such a large contribution is very small and connected to points where 
the Fermi surface touches the Brillouin zone (BZ) boundaries. A similar effect is known for 
the spin-mixing of Bloch states on the Fermi surface of Al.~\cite{Fabian_98,Gradhand09} 
The enhancement of the spin-mixing is induced by two mechanisms. Firstly, an avoided crossing of two bands appears at these points. 
Secondly, this avoided crossing occurs near the BZ boundary where the spin-orbit interaction is already increased due to the multiband character of the Fermi surface of Al.~\cite{Fabian_98} 
The same explanation, connected with the strength of the  
${\bf k}$-dependent spin-orbit coupling, holds for the enhancement of the Berry curvature. 
Except for these special points, the values for the Berry curvature in Al are 
orders of magnitude smaller than for all the other considered metals. In addition, 
the avoided crossings can explain the larger contributions in Pt in comparison to Au which 
is in fact heavier, but has only one degenerate band at the Fermi surface. Finally, Cu, also with one band only, 
has quite small contributions since it is a relatively light material. 

\section{Spin Hall conductivity}
As a first application of the Berry curvature, calculated within the KKR formalism, the intrinsic spin Hall conductivity (SHC) will be presented. 
This quantity was already calculated using a Kubo formula like expression for the SHE\cite{Yao05,Guo05,Freimuth_2010,Lowitzer_2011} and 
our purpose here is to validate our approach to the problem.

It might be preferable for such a comparison to calculate the anomalous Hall conductivity (AHC). 
As it is known, for this quantity the Kubo-like formula and the semiclassical expression are formally equivalent.\cite{Wang06} However, for nonmagnetic systems 
the AHC vanishes. This leaves us with no choice but to calculate the SHC in spite of two conceptual difficulties. 
The first of these is the lack of a proper definition of the spin current operator.\cite{Vernes_2007,Lowitzer_2010}
The second one is that, even with the frequently used choice of the spin current operator\cite{Guo05,Yao05,Sinova_2004}, the Kubo formula for the SHC 
is not equivalent to the simplified semiclassical theory used here.\cite{Shi2006}

In general, the AHC can be written in terms of the Berry curvature as\cite{KL_54,Luttinger_58,Sundaram_99,Thouless82,Yao04}
\begin{equation}\label{Eq.:AHC}
\sigma_{xy}=- \frac{e^2}{\hbar}\sum\limits_n\int_{BZ}\frac{d{\bf k}}{(2\pi)^3}f_n(E_F,{\bf k})\Omega_n^z({\bf k})\ ,
\end{equation}
where the distribution function $f_n(E_F,{\bf k})$ restricts the integral to the states below the Fermi energy $E_F$.

For the SHE this formula has to be modified to account for the fact that a spin and not a charge current is flowing. In addition, the non-Abelian 
nature of the Berry curvature has to be taken into account.\cite{Shindou05} 
Let us start with the heuristic spin-current operator $\underline{j}_s=1/2(\boldsymbol\Sigma \hat{\bf v}+\hat{\bf v}\boldsymbol\Sigma)$.\cite{Sinova_2004,Guo05}
Following the simplest interpretation of the semiclassical wave packet dynamics \cite{Sundaram_99,Culcer_2004} with 
$\hat{\bf v}\rightarrow\dot{\bf r}_c=-e\boldsymbol\Omega\times {\bf E}$ and $\boldsymbol\Sigma\rightarrow {\bf S}$, 
we consider the anomalous velocity induced by an applied electric field ${\bf E}$. Now, we have both 
${\bf S}$ and $\boldsymbol\Omega$ as $2\times 2$ 
matrices in the subspace spanned by the Kramers doublet. If we assume the electric field to be in $x$ direction and restrict the discussion to the 
spin polarization in $z$ direction, then the SHC is given by \cite{Shindou05}
\begin{equation}\label{Eq.:SHC}
\sigma^z_{xy}= \frac{e^2}{2\hbar}\sum\limits_n\int_{BZ}\frac{d{\bf k}}{(2\pi)^3}f_n(E_F,{\bf k})Tr[\rho_n({\bf k})S^n({\bf k})\Omega_n^z({\bf k})]\ .
\end{equation}
Here $\rho_n({\bf k})$ is the density matrix which describes the wave packet constructed from the two degenerate states corresponding to the wave vector ${\bf k}$ and 
band $n$. As mentioned above, this expression is not equivalent to the Kubo formula of Refs. \onlinecite{Wang07},~\onlinecite{Guo08},~\onlinecite{Freimuth_2010}, and 
\onlinecite{Lowitzer_2011}. The difference is induced by neglecting the band off-diagonal terms stemming from the spin operator.~\cite{Shindou05} 
Here we mean the other bands which are out of the considered Kramers doublet but may be energetically close to it. 
However, the Kramers doublet is treated correctly in terms of a non-Abelian Berry curvature.\cite{Shindou05} We leave a possible influence of these simplifications to be investigated elsewhere. 
Here we only show that within such approximations one can reproduce the results obtained in the more rigorous approach of the Kubo like formula.\cite{Yao05,Guo05,Guo08,Lowitzer_2011}
To aid the emergence of physical insight into the content of our calculations we made a further simplification by 
assuming the spin expectation value for the degenerate bands to be $S_{ii}^n({\bf k})=\pm 1$. 
This is equivalent to a two current model where the spin current is given by
$I^s = I^+ - I^-$. Here \lq\lq $+$\rq\rq\ and \lq\lq $-$\rq\rq\ denotes the current provided 
by $\Psi^+$ and $\Psi^-$ states with a positive or negative spin polarization, respectively.\cite{Gradhand09} 
Thus, the matrix element $\Omega_{n,11}^{z}$ in Eq.~(\ref{Eq.:SHC}) corresponds to $\Omega_n^{z,+}$. 
For an incoherent superposition of two wave packets corresponding to the degenerate states of the $n$th band 
the density matrix takes the form $\rho_n({\bf k})=\left(\begin{array}{cc}1&0\\0&1\end{array}\right)$. Therefore, the SHC can be written
as $\sigma_{xy}^z = \sigma_{xy}^+ - \sigma_{xy}^-$ leading to
\begin{eqnarray}
\sigma_{xy}^z &= & \frac{e^2}{\hbar} \sum\limits_n \int\limits_{BZ} \frac{d {\bf k}}{(2\pi)^3} f_n(E_F,{\bf k})\Omega_n^{z,+} ({\bf k})\label{Eq.:SHE_TwoCurr}\\
&=&  \frac{e^2}{\hbar(2\pi)^3}\int\limits^{E_F}d{\cal E}\Omega^z({\cal E})\nonumber\ ,
\end{eqnarray}
where
\begin{equation}
\Omega^z({\cal E})= \sum\limits_n\int\limits_{IS({\cal E})}\frac{d^2k}{\left|v^n_F({\bf k})\right|}\Omega_n^{z,+} ({\bf k})\ \text{.}\label{Eq.:SHE_TwoCurr_a}
\end{equation}
Here we exploited the fact that for the Kramers pair the condition $\Omega_n^{z,+}({\bf k})=-\Omega_n^{z,-}({\bf k})$ holds.
In fact, Eq.~(\ref{Eq.:SHE_TwoCurr}) is nothing else but the formula for the AHC applied for the
\lq\lq $+$\rq\rq\ subband only. It is written in terms of 
the energy-resolved Berry curvature $\Omega^z({\cal E})$ via an isosurface (IS) integral.

\begin{figure}[ht]
\includegraphics[width=0.96\linewidth]{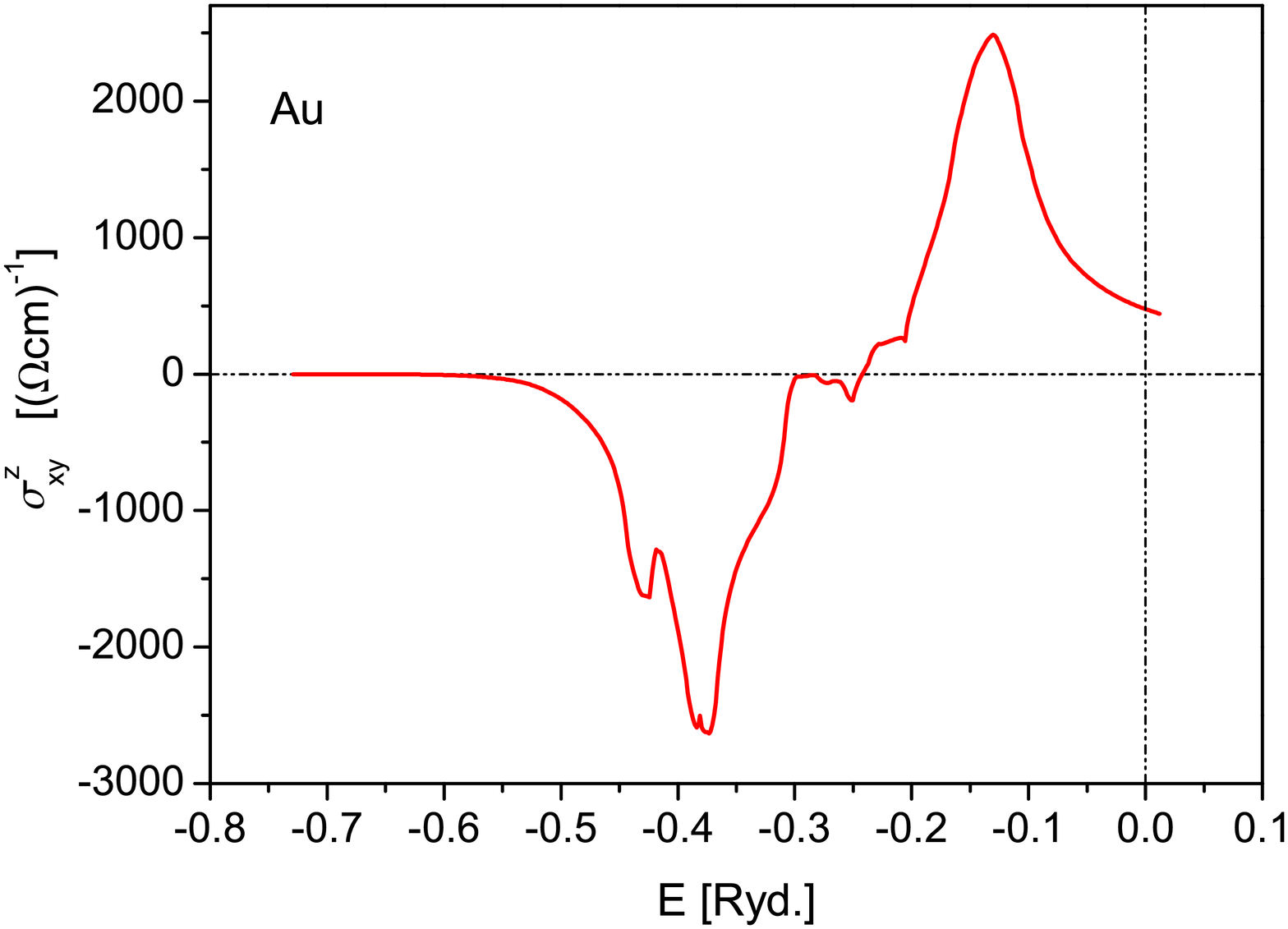}
\includegraphics[width=0.96\linewidth]{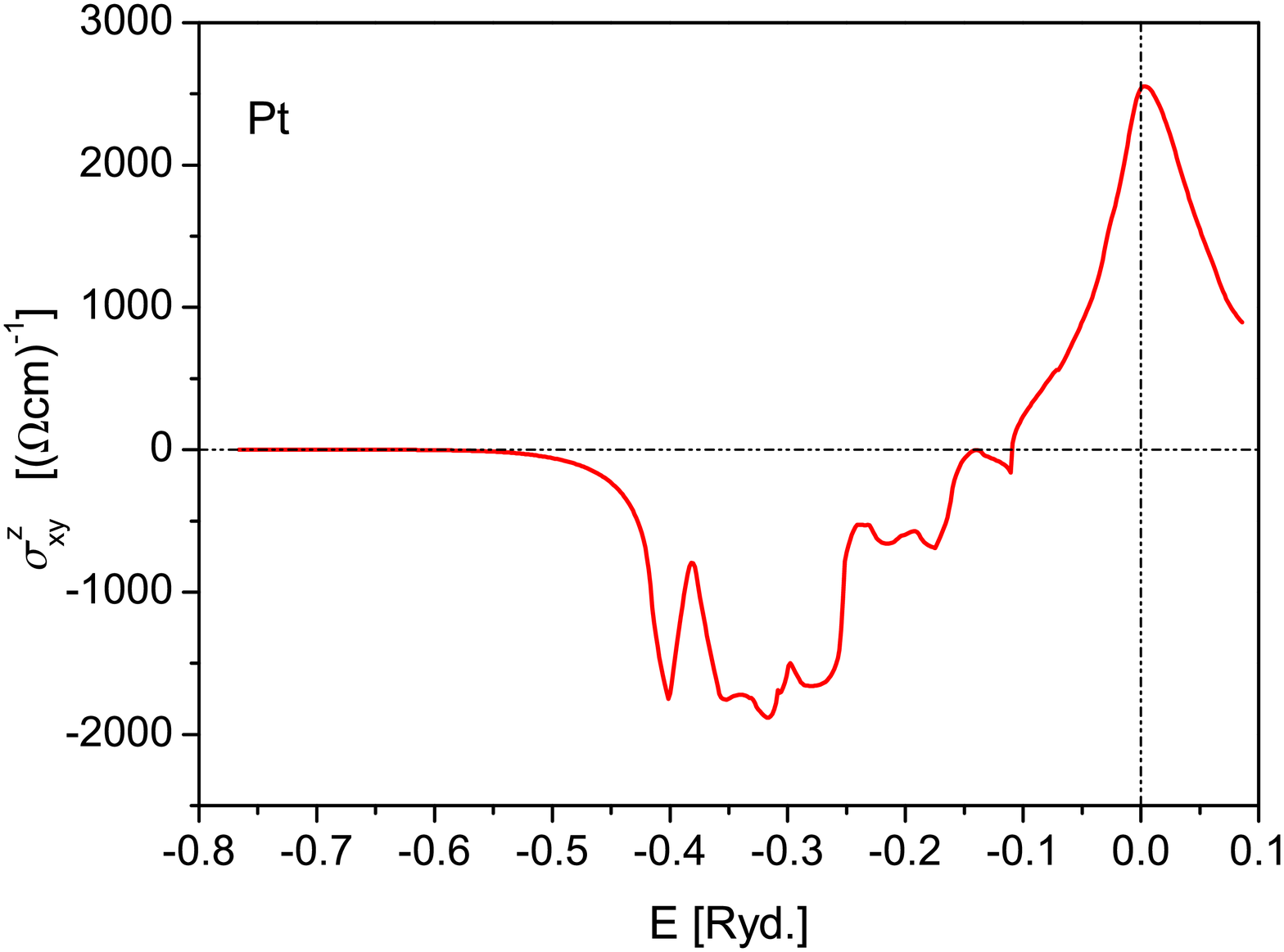}
\caption{The energy resolved spin Hall conductivity for Au (top) and Pt (bottom) according to Eq.~(\ref{Eq.:SHE_TwoCurr}).}\label{fig.:Au_sigma}
\end{figure}

In Fig. \ref{fig.:Au_sigma} we show the SHC as a function of $E_F$ for Au and Pt calculated by Eq. (\ref{Eq.:SHE_TwoCurr}). 
It is in reasonable agreement with the results obtained by Guo et al. \cite{Guo08APL,Guo08} using a Kubo formula approach. 
All main features in the energy dependence of the conductivity are reproduced. The conductivities at $E_F$ are given by $470\ (\Omega cm)^{-1}$ and $2500\ (\Omega cm)^{-1}$ 
for Au and Pt, respectively.

As it is well known, the integration of the Berry curvature over the Brillouin zone is a computationally very demanding task. \cite{Guo08,Yao05,Wang06} 
This stems from the fact that
$\boldsymbol\Omega ({\bf k})$ is a very spiky function in the crystal momentum space. Especially, for light elements the Berry curvature turns out to be small everywhere 
except for small regions around avoided crossings. The reason for that is already clear from the article of M.~Berry.~\cite{Berry84}. 
He expressed the curvature of a certain band as a sum over all the other bands where the difference of the band energies appear in the denominator. 
The same situation occurs in Eq.~(\ref{Curv_KKR_nonA}), where the eigenvalues of the KKR matrix play the role of the band energies.
Taking this into account, it is evident that the Berry curvature becomes larger if two bands are coming close to each other. 
This is exactly what happens at avoided crossings of any kind. As was pointed out by Mikitik and Sharlai in Ref.~\onlinecite{Mikitik1999}, the Berry curvature 
in the nonrelativistic case vanishes everywhere except for degeneracies of points or lines. In the vicinity of such degeneracies the Berry curvature is a $\delta$-distribution function.
Adding spin-orbit coupling to the system leads, normally, to avoided crossings at the degeneracies, but they still give rise to a Berry curvature. It can be viewed as smearing out 
the $\delta$ distribution. Importantly, the smearing is proportional to the strength of the spin-orbit coupling. 
It means that for light elements with a weak atomic spin-orbit coupling the Berry curvature is very close to the $\delta$ function. That makes the integration quite demanding. 
This leads to the somewhat surprising situation: systems with stronger spin-orbit coupling and more pronounced effects induced by the 
Berry curvature can be handled numerically easier than systems with tiny splitting of the bands.

\begin{figure}[ht]
\includegraphics[width=0.96\linewidth]{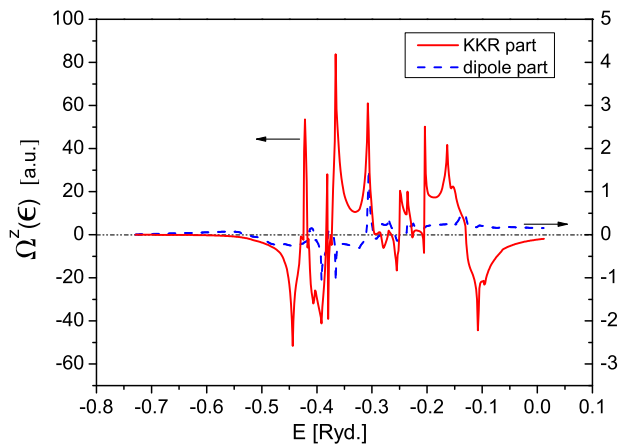}
\caption{The energy resolved Berry curvature for Au. The red (solid) and blue (dashed) curves show the separate contributions from the KKR part 
$\Omega^{KKR,z}_{11}({\cal E})$ and the dipole part 
$\Omega^{r,z}_{11}({\cal E})$ of the Berry curvature, respectively. }\label{fig.:Au_berry_E}
\end{figure}

To highlight once more the discussion above, 
in Fig. \ref{fig.:Au_berry_E} the energy-resolved Berry curvature according to Eq. (\ref{Eq.:SHE_TwoCurr_a}) is shown for Au.
Clearly, even with respect to energy 
$\Omega^z ({\cal E})$ is a very spiky function. That requires to use a very dense 
${\cal E}$ and ${\bf k}$ mesh as discussed by several authors. \cite{Guo08,Yao05,Wang06} 
Here we used comparable numbers of ${\bf k}$ points to converge the Berry curvature integrals. 
Actually, in Fig. \ref{fig.:Au_berry_E} only two of the three contributions, according to the separation given by Eqs. (\ref{Omega_ij})-(\ref{Curv_r_nonA}), are shown. 
One can see that the KKR part 
$\Omega^{KKR,z}_{11}({\cal E})$ (red solid line) dominates, whereas the part $\Omega^{r,z}_{11}({\cal E})$ (blue dashed lines) is negligible. 
We should mention that the part $\Omega^{v,z}_{11}({\cal E})$ is even much smaller and was skipped. 
This is a consequence of the above discussion related to Fig. \ref{fig.:berry_Au_all}.
Thus, only the most stable part of the Berry curvature, including no numerical derivative, contributes significantly to the SHC.

\section{Conclusion}

We have developed an efficient method to calculate the Berry curvature within the
KKR approach applied to the electronic structure of solids. An unconventional scheme that requires to deal with a
non-Hermitian KKR matrix is compensated by an elegant analytical
differentiation of this matrix with respect to the crystal momentum vector. This advantage is
a feature of the screened version of the KKR method and should also be useful for all tight-binding like computational methods. 
The formal arguments  starting with the local expansion of the Bloch function in Eq. (\ref{KKRexpansion}) and leading to the computable formulas 
(\ref{Curv_KKR_nonA})-(\ref{Curv_r_nonA}) can be readily adopted for calculations based on other multiple scattering approaches. 
In particular, for the LMTO method the situation will be simplified due to the lack of any energy dependence for the basis functions. 
The efficiency and stability of the proposed computational procedure is shown by calculating the Berry curvature for Al,
Cu, Au, and Pt bulk crystals and the spin Hall conductivity for Au and Pt.

\appendix

\section{Derivation of the group velocity}\label{App_a}

Following Shilkova and Shirokovskii \cite{Shilkova88} we note that the eigenvalues of the KKR-matrix obey
$\lambda_n ({\cal E}_n({\bf k}); {\bf k})= 0$ and hence for the total derivative
$\boldsymbol\nabla_{\bf k} \lambda_n ({\cal E}; {\bf k})$ we have
\begin{equation}
\begin{array}{ll}
\frac{\partial \lambda_n ({\cal E}; {\bf k})}{\partial {\bf k}} |_
{_{_{{\cal E}={\cal E}_n ({\bf k})}}} + \boldsymbol\nabla_{\bf k} {\cal E}_n ({\bf k})
\frac{\partial \lambda_n ({\cal E}; {\bf k})}{\partial {\cal E}}|_
{_{_{{\cal E}={\cal E}_n ({\bf k})}}} = 0\ .
\end{array}
\label{eqno(8)}
\end{equation}
Therefore,
\begin{equation}
\begin{array}{ll}
{\bf v}_n ({\bf k}) = \boldsymbol\nabla_{\bf k} {\cal E}_n ({\bf k}) =
- \frac{\partial \lambda _{n} ({\cal E}; {\bf k})}{\partial {\bf k}} / 
\frac{\partial \lambda _{n} ({\cal E}; {\bf k})}
{\partial {\cal E}}\ . 
\end{array}
\label{eqno(8.a)}
\end{equation}

The next and central move is to calculate $\partial \lambda _{n} 
({\cal E}; {\bf k}) / \partial {\bf k}$. For the case of Hermitian
KKR matrix $\Bar{\Bar M} ({\cal E}; {\bf k})$ it was done in Ref.~\onlinecite{Shilkova88}.
Here we generalize the procedure to the case of a non-Hermitian matrix. 
With the definition of the left eigenvectors \cite{Kalaba81} 
${\bar D}_n^\dagger \Bar{\Bar M} = \lambda_n {\bar D}_n^\dagger$ it is evident that 
\begin{equation}
\begin{array}{ll}
{\bar D}_n^\dagger \Bar{\Bar M} {\bar C}_n = \lambda_n
{\bar D}_n^\dagger {\bar C}_n\ . 
\end{array}
\label{eqno(9)}
\end{equation}
Taking the partial derivative of both sides
of Eq.~(\ref{eqno(9)}) with respect to ${\bf k}$ we obtain
\begin{equation}  
\begin{array}{ll}
\frac{\partial \lambda _{n} ({\cal E}; {\bf k})}{\partial {\bf k}} =
\frac{D_n^\dagger (\partial \Bar{\Bar M} / \partial {\bf k}) C_n}
{D_n^\dagger C_n}\ .
\end{array}
\end{equation}
Using this formula in Eq.~(\ref{eqno(8.a)}) one derives Eq.~(\ref{eqno8c}).

\section{Connection 
${\bf \cal A}_n^k({\bf k})$}\label{App_b}

To deal with Eq.~(\ref{localconnection}), we need to calculate $\boldsymbol\nabla_{\bf k} 
\Psi_{n {\bf k}} ({\bf r})$. Using the KKR expansion of Eq.~(\ref{KKRexpansion}), we obtain 
\begin{equation}
\begin{array}{ll}
\boldsymbol\nabla_{\bf k} \Psi_{n {\bf k}} ({\bf r}) =
\sum\limits_Q [ \boldsymbol\nabla_{\bf k} C_Q^n ({\bf k}) \Phi_Q ({\bf r}) +
C_Q^n ({\bf k}) {\bf v}_n ({\bf k}) \frac{\partial \Phi_Q ({\bf r})}{\partial {\cal E}}]
\end{array}
\end{equation}
that gives us
\begin{widetext}
\begin{equation}
\begin{array}{ll}
\langle \Psi_{n {\bf k}} | \boldsymbol\nabla_{\bf k} \Psi_{n {\bf k}} \rangle =
\sum\limits_Q C_Q^{n *} ({\bf k}) \boldsymbol\nabla_{\bf k} C_Q^n ({\bf k}) + 
{\bf v}_n ({\bf k})\sum\limits_Q |C_Q^n ({\bf k})|^2 \int\limits_w \Phi_Q^\dagger 
({\cal E}; {\bf r}) \frac{\partial \Phi_Q ({\cal E}; {\bf r})}{\partial {\cal E}} d {\bf r} \ .
\end{array}
\end{equation}
Here we have used the fact that $\int_w \Phi_Q^\dagger ({\cal E}; {\bf r}) 
\frac{\partial \Phi_{Q^\prime} ({\cal E}; {\bf r})}{\partial {\cal E}} d {\bf r}
\propto\delta_{Q Q^\prime}$ by the properties of our KKR-basis set given by 
Eq.~(\ref{basis_set}). Then for ${\bf \cal A}_n^k({\bf k})$ we can write
\begin{equation}
\begin{array}{ll}
{\bf \cal A}_n^k({\bf k}) = {\bf \cal A}_n^{KKR}({\bf k}) +  
i\ {\bf v}_n ({\bf k})\sum\limits_Q |C_Q^n ({\bf k})|^2 
\int\limits_w \Phi_Q^\dagger ({\cal E}; {\bf r}) 
\frac{\partial \Phi_Q ({\cal E}; {\bf r})}{\partial {\cal E}} d {\bf r}\ ,\ \ 
\text{where}\ \ 
{\bf \cal A}_n^{KKR}({\bf k}) = i \sum\limits_Q C_Q^{n *} ({\bf k}) 
\boldsymbol\nabla_{\bf k} C_Q^n ({\bf k})\ . 
\end{array}
\label{Ak_QQ}
\end{equation}
Rewriting these expressions in a matrix form, we get Eqs.~(3.3)--(3.6).

\section{Abelian curvature}\label{App_c}

Let us derive first Eq.~(\ref{KKRomega}) for $\boldsymbol\Omega_n^{KKR} ({\bf k})$ starting 
from Eq.~(\ref{KKRcurvatureB}). Using the completeness relation given by 
Eq.~(\ref{completeness}) we can perform the following expansion
\begin{equation}
\begin{array}{ll}
{\bar C}_n = \sum\limits_m \frac{{\bar D}_m {\bar C}_m^\dagger}{{\bar C}_m^\dagger {\bar D}_m}
{\bar C}_n = \sum\limits_m \frac{{\bar C}_m^\dagger {\bar C}_n}{{\bar C}_m^\dagger {\bar D}_m}
{\bar D}_m\ .
\end{array}
\label{C_expansion}
\end{equation}
With the Hermitian conjugate of this expansion we have
\begin{equation}
\begin{array}{ll}
{\bar C}_n^\dagger \boldsymbol\nabla_{\bf k} {\bar C}_n = 
\frac{{\bar D}_n^\dagger \boldsymbol\nabla_{\bf k} {\bar C}_n}{{\bar D}_n^\dagger {\bar C}_n} +
\sum\limits_{m \ne n} \frac{{\bar C}_n^\dagger {\bar C}_m}{{\bar D}_m^\dagger {\bar C}_m}
{\bar D}_m^\dagger \boldsymbol\nabla_{\bf k} {\bar C}_n\ .
\end{array}
\label{CC_DC}
\end{equation}
Then, 
\begin{equation}
\begin{array}{rr}
\frac{\boldsymbol\nabla_{\bf k} {\bar C}_n^\dagger {\bar C}_n \times
{\bar D}_n^\dagger \boldsymbol\nabla_{\bf k} {\bar C}_n }{{\bar D}_n^\dagger {\bar C}_n} + 
\sum\limits_{m \ne n} \frac{\boldsymbol\nabla_{\bf k} {\bar C}_n^\dagger {\bar C}_m \times
{\bar D}_m^\dagger \boldsymbol\nabla_{\bf k} {\bar C}_n }{{\bar D}_m^\dagger {\bar C}_m} 
= \sum\limits_{m \ne n} \frac{ \boldsymbol\nabla_{\bf k} {\bar C}_n^\dagger [{\bar C}_m -
{\bar C}_n ({\bar C}_n^\dagger {\bar C}_m)] \times {\bar D}_m^\dagger \boldsymbol\nabla_{\bf k} 
{\bar C}_n} {{\bar D}_m^\dagger {\bar C}_m}\ ,
\end{array}
\label{equation7.11}
\end{equation}
where we have used that $\boldsymbol\nabla_{\bf k} {\bar C}_n^\dagger {\bar C}_n \times 
{\bar C}_n^\dagger \boldsymbol\nabla_{\bf k} {\bar C}_n$ vanishes since 
${\bar C}_n^\dagger \boldsymbol\nabla_{\bf k} {\bar C}_n$ is purely imaginary. 
Using Eq.~(\ref{C_expansion}) we can rewrite the part of the numerator in 
Eq.~(\ref{equation7.11}) as
\begin{equation}
\begin{array}{rr}
{\bar C}_m - {\bar C}_n ({\bar C}_n^\dagger {\bar C}_m) = 
\sum\limits_k \frac{{\bar C}_k^\dagger {\bar C}_m}{{\bar C}_k^\dagger {\bar D}_k} {\bar D}_k
- \sum\limits_k \frac{({\bar C}_k^\dagger {\bar C}_n)({\bar C}_n^\dagger {\bar C}_m)}
{{\bar C}_k^\dagger {\bar D}_k} {\bar D}_k = \sum\limits_{k \ne n}
\frac{[{\bar C}_k^\dagger {\bar C}_m - ({\bar C}_k^\dagger {\bar C}_n)({\bar C}_n^\dagger 
{\bar C}_m) ] {\bar D}_k}{{\bar C}_k^\dagger {\bar D}_k}\ . 
\end{array}
\end{equation}
In addition, due to ${\bar D}_j^\dagger {\bar C}_i =0$ for $j \ne i$, we have
\begin{equation}
{\bar D}_j^\dagger \boldsymbol\nabla_{\bf k} {\bar C}_i =
\frac{{\bar D}_j^\dagger \boldsymbol\nabla_{\bf k} \Bar{\Bar M} {\bar C}_i}
{\lambda_i - \lambda_j}\ \ ,\ \ \ j \ne i\ \ .
\label{Dj_nabla_Ci}
\end{equation}
Therefore, finally we can write
\begin{equation}
\begin{array}{ll}
\sum\limits_m \frac{\boldsymbol\nabla_{\bf k} {\bar C}_n^\dagger {\bar C}_m \times
{\bar D}_m^\dagger \boldsymbol\nabla_{\bf k} {\bar C}_n }{{\bar D}_m^\dagger {\bar C}_m} 
= \sum\limits_{m \ne n} \frac 1{{\bar D}_m^\dagger {\bar C}_m
(\lambda_n - \lambda_m)} \sum\limits_{k \ne n} 
\frac{[{\bar C}_k^\dagger {\bar C}_m - ({\bar C}_k^\dagger {\bar C}_n) 
({\bar C}_n^\dagger {\bar C}_m) ]}{{\bar C}_k^\dagger {\bar D}_k} 
\frac{({\bar D}_k^\dagger \boldsymbol\nabla_{\bf k} \Bar{\Bar M} 
{\bar C}_n)^* \times {\bar D}_m^\dagger \boldsymbol\nabla_{\bf k} 
\Bar{\Bar M} {\bar C}_n} { (\lambda_n^* - \lambda_k^*)}
\end{array}
\end{equation}
and end up with Eq.~(\ref{KKRomega}). 

For a derivation of $\boldsymbol\Omega_n^v ({\bf k})$ we need to take the curl of the second term 
in r.h.s. of Eq.~(\ref{Ak_QQ}). Taking into account that 
$\boldsymbol\nabla_{\bf k} \times {\bf v}_n ({\bf k}) = \boldsymbol\nabla_{\bf k} \times
\boldsymbol\nabla_{\bf k} {\cal E}_n ({\bf k}) =0$, one can write
\begin{equation}
\begin{array}{ll}
\boldsymbol\Omega_n^v ({\bf k}) = 2 {\bf v}_n ({\bf k}) \times 
Im \{ \sum\limits_Q [C_Q^{n *} ({\bf k}) \boldsymbol\nabla_{\bf k} C_Q^n ({\bf k})] 
\int\limits_w \Phi_Q ({\cal E}; {\bf r}) 
\frac{\partial \Phi_Q^\dagger ({\cal E}; {\bf r})}{\partial {\cal E}} d {\bf r} \}
= 2 {\bf v}_n ({\bf k}) \times 
Im \{ {\bar C}_n^\dagger \Bar{\Bar \Delta} \boldsymbol\nabla_{\bf k} {\bar C}_n \}\ .
\end{array}
\end{equation}
Then, using again the completeness relation of Eq.~(\ref{completeness}) together with 
Eqs.~(\ref{CC_DC}) and (\ref{Dj_nabla_Ci}) we obtain
\begin{equation}
\begin{array}{rr}
{\bar C}_n^\dagger \Bar{\Bar \Delta} \boldsymbol\nabla_{\bf k} {\bar C}_n =
\sum\limits_m \frac{{\bar C}_n^\dagger \Bar{\Bar \Delta} {\bar C}_m {\bar D}_m^\dagger
\boldsymbol\nabla_{\bf k} {\bar C}_n}{{\bar D}_m^\dagger {\bar C}_m} =
{\bar C}_n^\dagger \Bar{\Bar \Delta} {\bar C}_n
{\bar C}_n^\dagger \boldsymbol\nabla_{\bf k} {\bar C}_n +  
\sum\limits_{m \ne n} \left[ \frac{ {\bar C}_n^\dagger \Bar{\Bar \Delta} {\bar C}_m
{\bar D}_m^\dagger \boldsymbol\nabla_{\bf k} \Bar{\Bar M} {\bar C}_n}
{{\bar D}_m^\dagger {\bar C}_m (\lambda_n - \lambda_m)} -
\frac{ ({\bar C}_n^\dagger \Bar{\Bar \Delta} {\bar C}_n)
{\bar D}_m^\dagger \boldsymbol\nabla_{\bf k} \Bar{\Bar M} {\bar C}_n 
({\bar C}_n^\dagger {\bar C}_m)} {{\bar D}_m^\dagger {\bar C}_m (\lambda_n - \lambda_m)}
\right]\ . 
\end{array}
\end{equation}
Here the term ${\bar C}_n^\dagger \Bar{\Bar \Delta} {\bar C}_n
{\bar C}_n^\dagger \boldsymbol\nabla_{\bf k} {\bar C}_n$ is purely real since 
${\bar C}_n^\dagger \boldsymbol\nabla_{\bf k} {\bar C}_n$ and 
${\bar C}_n^\dagger \Bar{\Bar \Delta} {\bar C}_n$ both are purely imaginary
quantities (latter one due to the normalization $\int_w |\Phi_Q|^2 d {\bf r} =1$).
Hence we end up with Eq.~(\ref{Omegav}). 

Let us consider now Eq.~(\ref{rcurvature}) and use the KKR expansion given by 
Eq.~(\ref{KKRexpansion}). Then, 
\begin{equation}
\begin{array}{ll}
\boldsymbol\Omega_n^r ({\bf k}) = 2 Re \{ \sum\limits_{Q Q^\prime} \int\limits_\omega 
[ {\bar C}_Q^{n *}  \Phi_Q^\dagger \boldsymbol\nabla_{\bf k} ({\bar C}_{Q^\prime}^n 
\Phi_{Q^\prime}) \times {\bf r} ] d {\bf r} \} = 
2 Re \{ \sum\limits_{Q Q^\prime} \int\limits_\omega [ {\bf v}_n ({\bf k}) \times
{\bar C}_Q^{n *}  \Phi_Q^\dagger {\bf r} \frac{\partial \Phi_{Q^\prime} ({\cal E};{\bf r})}
{\partial {\cal E}} {\bar C}_{Q^\prime}^n ] d {\bf r} \} - \\ \\ -
2 Re \{ \sum\limits_{Q Q^\prime} \int\limits_\omega [ {\bar C}_Q^{n *} \Phi_Q^\dagger {\bf r} 
\Phi_{Q^\prime} \times \boldsymbol\nabla_{\bf k} {\bar C}_{Q^\prime}^n ] d {\bf r} \}
= 2 Re \{ {\bf v}_n ({\bf k}) \times 
{\bar C}_n^\dagger {\Bar {\Bar {\bf r}}}_{\cal E} {\bar C}_n \} - 2 Re \{ {\bar C}_n^\dagger 
{\Bar {\Bar {\bf r}}} \times \boldsymbol\nabla_{\bf k} {\bar C}_n \}\ .
\end{array}
\label{Omrmat}
\end{equation}  
Using again the completeness relation of Eq.~(\ref{completeness}) together with Eqs.~(\ref{CC_DC}) 
and (\ref{Dj_nabla_Ci}), for the second term of Eq.~(\ref{Omrmat}) we can write 
\begin{equation}
\begin{array}{ll}
{\bar C}_n^\dagger {\Bar {\Bar {\bf r}}} \times \boldsymbol\nabla_{\bf k} {\bar C}_n =
\sum\limits_m \frac{{\bar C}_n^\dagger {\Bar {\Bar {\bf r}}} {\bar C}_m \times 
{\bar D}_m^\dagger \boldsymbol\nabla_{\bf k} {\bar C}_n} {{\bar D}_m^\dagger {\bar C}_m} = 
{\bar C}_n^\dagger {\Bar {\Bar {\bf r}}} {\bar C}_n \times {\bar C}_n^\dagger 
\boldsymbol\nabla_{\bf k} {\bar C}_n - \sum\limits_{m \ne n} 
\frac{ {\bar C}_n^\dagger {\Bar {\Bar {\bf r}}} {\bar C}_n \times
{\bar D}_m^\dagger \boldsymbol\nabla_{\bf k} \Bar{\Bar M} 
{\bar C}_n ({\bar C}_n^\dagger {\bar C}_m)}{{\bar D}_m^\dagger {\bar C}_m 
(\lambda_n - \lambda_m)} + \sum\limits_{m \ne n} \frac{ {\bar C}_n^\dagger {\Bar {\Bar {\bf r}}} 
{\bar C}_m \times {\bar D}_m^\dagger \boldsymbol\nabla_{\bf k} \Bar{\Bar M} {\bar C}_n}
{{\bar D}_m^\dagger {\bar C}_m (\lambda_n - \lambda_m)}\ .
\end{array}
\end{equation}  
Here ${\bar C}_n^\dagger {\Bar {\Bar {\bf r}}} {\bar C}_n \times {\bar C}_n^\dagger 
\boldsymbol\nabla_{\bf k} {\bar C}_n$ does not contribute to Eq.~(\ref{Omrmat})
since the quantity ${\bar C}_n^\dagger {\Bar {\Bar {\bf r}}} {\bar C}_n$ is purely real while 
${\bar C}_n^\dagger \boldsymbol\nabla_{\bf k} {\bar C}_n$ is purely imaginary. Thus, finally we obtain 
Eq.~(\ref{rmatcurvature}).

\section{Non-Abelian curvature}\label{App_d}

We start with $\boldsymbol\Omega_{ij}^k ({\bf k})$ part of the representation for the
non-Abelian curvature given by Eq.~(\ref{Omega_ij}). The first term contributing to this
part is
\begin{equation}
\begin{array}{ll}
\boldsymbol\Omega_{ij}^{KKR} ({\bf k}) = i \boldsymbol\nabla_{\bf k} {\bar C}_i^\dagger
\times \boldsymbol\nabla_{\bf k} {\bar C}_j - i \sum\limits_{l \in \Sigma} 
\boldsymbol\nabla_{\bf k} {\bar C}_i^\dagger
{\bar C}_l \times {\bar C}_l^\dagger \boldsymbol\nabla_{\bf k} {\bar C}_j = 
i \sum\limits_m \frac{\boldsymbol\nabla_{\bf k} {\bar C}_i^\dagger {\bar C}_m \times
{\bar D}_m^\dagger \boldsymbol\nabla_{\bf k} {\bar C}_j}{{\bar D}_m^\dagger {\bar C}_m} - \\ 
\\ - i \sum\limits_{l \in \Sigma} \boldsymbol\nabla_{\bf k} {\bar C}_i^\dagger {\bar C}_l \times
\left[ \frac {{\bar D}_l^\dagger} {{\bar D}_l^\dagger {\bar C}_l} +
\sum\limits_{m \notin \Sigma} \frac {{\bar C}_l^\dagger {\bar C}_m} 
{{\bar D}_m^\dagger {\bar C}_m}  {\bar D}_m^\dagger \right] \boldsymbol\nabla_{\bf k} {\bar C}_j =
i \sum\limits_{m \notin \Sigma}  
\frac{[ \boldsymbol\nabla_{\bf k} {\bar C}_i^\dagger {\bar C}_m - \sum_{l \in \Sigma}
\boldsymbol\nabla_{\bf k} {\bar C}_i^\dagger {\bar C}_l 
({\bar C}_l^\dagger {\bar C}_m)] \times {\bar D}_m^\dagger \boldsymbol\nabla_{\bf k} 
\Bar{\Bar M} {\bar C}_j} {{\bar D}_m^\dagger {\bar C}_m (\lambda_j - \lambda_m)}\ ,
\end{array}
\end{equation}
where we have used Eqs.~(\ref{C_expansion}) and (\ref{Dj_nabla_Ci}). 
According to Eq.~(\ref{C_expansion}), we can rewrite the term in the square brackets as
\begin{equation}
\begin{array}{ll}
\boldsymbol\nabla_{\bf k} {\bar C}_i^\dagger \{ {\bar C}_m - \sum\limits_{l \in \Sigma}
{\bar C}_l ({\bar C}_l^\dagger {\bar C}_m) \} =
\boldsymbol\nabla_{\bf k} {\bar C}_i^\dagger \{ \sum\limits_k 
\frac{{\bar C}_k^\dagger {\bar C}_m}{{\bar C}_k^\dagger {\bar D}_k}{\bar D}_k - 
\sum\limits_{l \in \Sigma} ({\bar C}_l^\dagger {\bar C}_m) [ \frac{{\bar D}_l}
{{\bar C}_l^\dagger {\bar D}_l} + \sum\limits_{k \notin \Sigma} \frac{{\bar C}_k^\dagger 
{\bar C}_l}{{\bar C}_k^\dagger {\bar D}_k}{\bar D}_k ] \} = \\ \\
\ \ \ \ \ \ \ \ \ \ \ \ \ \ \ \ \  
= \sum\limits_{k \notin \Sigma} \frac{\boldsymbol\nabla_{\bf k} {\bar C}_i^\dagger {\bar D}_k 
[{\bar C}_k^\dagger {\bar C}_m - \sum_{l \in \Sigma} ({\bar C}_k^\dagger {\bar C}_l)
({\bar C}_l^\dagger {\bar C}_m)]}{{\bar C}_k^\dagger {\bar D}_k} =
\sum\limits_{k \notin \Sigma}  \frac{ [{\bar C}_k^\dagger {\bar C}_m -
\sum_{l \in \Sigma} 
({\bar C}_k^\dagger {\bar C}_l)({\bar C}_l^\dagger {\bar C}_m)] ({\bar D}_k^\dagger
\boldsymbol\nabla_{\bf k} \Bar{\Bar M} {\bar C}_i)^* }
{{\bar C}_k^\dagger {\bar D}_k (\lambda_i^* - \lambda_k^*)}\ .
\end{array}
\end{equation}
Therefore, we end up with Eq.~(\ref{Curv_KKR_nonA}).
Now we consider the second term contributing to $\boldsymbol\Omega_{ij}^k ({\bf k})$. Namely,
\begin{equation}
\begin{array}{lll}
\boldsymbol\Omega_{ij}^v ({\bf k}) &=& i [ {\bf v}_i \times {\bar C}_i^\dagger
\Bar{\Bar \Delta}^\dagger \boldsymbol\nabla_{\bf k} {\bar C}_j - {\bf v}_j \times 
\boldsymbol\nabla_{\bf k} {\bar C}_i^\dagger \Bar{\Bar \Delta} {\bar C}_j ] -   
i \sum\limits_{l \in \Sigma} \{
{\bf v}_i {\bar C}_i^\dagger \Bar{\Bar \Delta}^\dagger {\bar C}_l \times {\bar C}_l^\dagger
\boldsymbol\nabla_{\bf k} {\bar C}_j - {\bf v}_j \times \boldsymbol\nabla_{\bf k} 
{\bar C}_i^\dagger {\bar C}_l ({\bar C}_l^\dagger \Bar{\Bar \Delta} {\bar C}_j) \}\ +\\
&+&i\left[{\bf v}_i\times {\bf v}_j\right]\left\{\bar{c}_i^{\dagger}\bar{\bar \Delta}_{\cal E} 
\bar{c}_j-\sum\limits_{l\in \Sigma}(\bar{c}_i^{\dagger}\bar{\bar \Delta}^{\dagger} \bar{c}_l)(\bar{c}_l^{\dagger}\bar{\bar \Delta} \bar{c}_j)\right\}\ ,
\end{array}
\end{equation}
where the matrix $\bar{\bar \Delta}_{\cal E} $ is defined by Eq.~(\ref{DeltaEmatrix}).
Here, due to Eqs.~(\ref{completeness}) and (\ref{C_expansion}), we have
\begin{equation}
\begin{array}{ll}
{\bf v}_i \times {\bar C}_i^\dagger \Bar{\Bar \Delta}^\dagger \boldsymbol\nabla_{\bf k} 
{\bar C}_j = {\bf v}_i \times \sum\limits_m \frac{{\bar C}_i^\dagger
\Bar{\Bar \Delta}^\dagger {\bar C}_m {\bar D}_m^\dagger \boldsymbol\nabla_{\bf k} {\bar C}_j}
{{\bar D}_m^\dagger {\bar C}_m}\ ,\ \ \  
{\bf v}_j \times \boldsymbol\nabla_{\bf k} {\bar C}_i^\dagger \Bar{\Bar \Delta} {\bar C}_j =
 {\bf v}_j \times \sum\limits_m \frac{ \boldsymbol\nabla_{\bf k} {\bar C}_i^\dagger {\bar D}_m
{\bar C}_m^\dagger \Bar{\Bar \Delta} {\bar C}_j}
{{\bar C}_m^\dagger {\bar D}_m}
\ ,\\  
\sum\limits_{l \in \Sigma} {\bf v}_i {\bar C}_i^\dagger \Bar{\Bar \Delta}^\dagger {\bar C}_l 
\times {\bar C}_l^\dagger \boldsymbol\nabla_{\bf k} {\bar C}_j = {\bf v}_i \times 
\sum\limits_{l \in \Sigma} {\bar C}_i^\dagger \Bar{\Bar \Delta}^\dagger {\bar C}_l 
[ \frac{{\bar D}_l^\dagger} {{\bar D}_l^\dagger {\bar C}_l} + \sum\limits_{m \notin \Sigma} 
\frac{{\bar C}_l^\dagger {\bar C}_m}{{\bar D}_m^\dagger {\bar C}_m} 
{\bar D}_m^\dagger ] \boldsymbol\nabla_{\bf k} {\bar C}_j
\ ,\\ 
\sum\limits_{l \in \Sigma} {\bf v}_j \times \boldsymbol\nabla_{\bf k} 
{\bar C}_i^\dagger {\bar C}_l ({\bar C}_l^\dagger \Bar{\Bar \Delta} {\bar C}_j) = 
{\bf v}_j \times \sum\limits_{l \in \Sigma}
({\bar C}_l^\dagger \Bar{\Bar \Delta} {\bar C}_j) \boldsymbol\nabla_{\bf k} {\bar C}_i^\dagger
[ \frac{{\bar D}_l}{{\bar C}_l^\dagger {\bar D}_l} + 
\sum\limits_{m \notin \Sigma} \frac{{\bar C}_m^\dagger {\bar C}_l}{{\bar C}_m^\dagger {\bar D}_m} 
{\bar D}_m ] \ .
\end{array}
\end{equation}
Hence we end up with Eq.~(\ref{Curv_v_nonA}). Let us consider now 
\begin{equation}
\begin{array}{ll}
\boldsymbol\Omega_{ij}^r ({\bf k}) = \langle \boldsymbol\nabla_{\bf k} \Psi_i \times {\bf r} |  
\Psi_j \rangle - \langle \Psi_i | {\bf r} \times \boldsymbol\nabla_{\bf k} \Psi_j \rangle + \\
\\ + \sum\limits_{l \in \Sigma} \left\{
\langle \Psi_i |{\bf r}| \Psi_l \rangle \times \langle \Psi_l | \boldsymbol\nabla_{\bf k}
\Psi_j \rangle -  \langle \boldsymbol\nabla_{\bf k} \Psi_i | 
\Psi_l \rangle \times \langle \Psi_l |{\bf r}| \Psi_j \rangle -
i \langle \Psi_i |{\bf r}| \Psi_l \rangle \times \langle \Psi_l |{\bf r}| \Psi_j \rangle 
\right\}\ . 
\end{array}
\label{Omegaij_r}
\end{equation}
Here, due to the completeness relation given by Eq.~(\ref{completeness}),
\begin{equation}
\begin{array}{ll}
\langle \boldsymbol\nabla_{\bf k} \Psi_i \times {\bf r} | \Psi_j \rangle = 
{\bf v}_i \times {\bar C}_i^\dagger {\Bar {\Bar {\bf r}}}_{\cal E} {\bar C}_j +
\boldsymbol\nabla_{\bf k} {\bar C}_i^\dagger \times {\Bar{\Bar {\bf r}}} {\bar C}_j 
= {\bf v}_i \times {\bar C}_i^\dagger {\Bar {\Bar {\bf r}}}_{\cal E} {\bar C}_j +
\sum\limits_m \frac{\boldsymbol\nabla_{\bf k} {\bar C}_i^\dagger {\bar D}_m \times 
{\bar C}_m^\dagger {\Bar{\Bar {\bf r}}} {\bar C}_j}{{\bar C}_m^\dagger {\bar D}_m}\ , \\
\langle \Psi_i | {\bf r} \times \boldsymbol\nabla_{\bf k} \Psi_j \rangle = 
- {\bf v}_j \times {\bar C}_i^\dagger {\Bar {\Bar {\bf r}}}_{\cal E} {\bar C}_j +
{\bar C}_i^\dagger {\Bar{\Bar {\bf r}}} \times \boldsymbol\nabla_{\bf k} {\bar C}_j = 
- {\bf v}_j \times {\bar C}_i^\dagger {\Bar {\Bar {\bf r}}}_{\cal E} {\bar C}_j +
\sum\limits_m \frac{{\bar C}_i^\dagger {\Bar{\Bar {\bf r}}} {\bar C}_m \times
{\bar D}_m^\dagger \boldsymbol\nabla_{\bf k} {\bar C}_j}{{\bar D}_m^\dagger {\bar C}_m}\ . 
\end{array}
\end{equation}
In addition, taking into account Eq.~(\ref{C_expansion}), we have
\begin{equation}
\begin{array}{ll}
\sum\limits_{l \in \Sigma} \langle \Psi_i |{\bf r}| \Psi_l \rangle \times \langle \Psi_l | 
\boldsymbol\nabla_{\bf k} \Psi_j \rangle = \sum\limits_{l \in \Sigma} {\bar C}_i^\dagger 
{\Bar{\Bar {\bf r}}} {\bar C}_l \times [{\bar C}_l^\dagger \boldsymbol\nabla_{\bf k} {\bar C}_j 
+ {\bf v}_j {\bar C}_l^\dagger \Bar{\Bar \Delta} {\bar C}_j] = \\   
\ \ \ \ \ \ \ \ \ \ \ \ \ \ \ \ \ \ \ \ \ \ \ \ \ \ \ \ \ \ \ \ \ \ 
= \sum\limits_{l \in \Sigma} \{ {\bar C}_i^\dagger {\Bar{\Bar {\bf r}}} {\bar C}_l \times
[ \frac{{\bar D}_l^\dagger}{{\bar D}_l^\dagger {\bar C}_l} + \sum\limits_{m \notin \Sigma}
\frac{{\bar C}_l^\dagger {\bar C}_m}{{\bar D}_m^\dagger {\bar C}_m} {\bar D}_m^\dagger ] 
\boldsymbol\nabla_{\bf k} {\bar C}_j - {\bf v}_j \times
({\bar C}_i^\dagger {\Bar{\Bar {\bf r}}} {\bar C}_l) 
({\bar C}_l^\dagger \Bar{\Bar \Delta} {\bar C}_j) \}\ , \\
\sum\limits_{l \in \Sigma} \langle \boldsymbol\nabla_{\bf k} \Psi_i | 
\Psi_l \rangle \times \langle \Psi_l |{\bf r}| \Psi_j \rangle = \sum\limits_{l \in \Sigma} 
[\boldsymbol\nabla_{\bf k} {\bar C}_i^\dagger {\bar C}_l + {\bf v}_i 
{\bar C}_i^\dagger \Bar{\Bar \Delta}^\dagger {\bar C}_i ] \times {\bar C}_l^\dagger 
{\Bar{\Bar {\bf r}}} {\bar C}_j = \\
\ \ \ \ \ \ \ \ \ \ \ \ \ \ \ \ \ \ \ \ \ \ \ \ \ \ \ \ \ \ \ \ \ \ 
= \sum\limits_{l \in \Sigma} \{
\boldsymbol\nabla_{\bf k} {\bar C}_i^\dagger [ \frac{{\bar D}_l}{{\bar C}_l^\dagger 
{\bar D}_l} + \sum\limits_{m \notin \Sigma} \frac{{\bar C}_m^\dagger {\bar C}_l}
{{\bar C}_m^\dagger {\bar D}_m} {\bar D}_m ] \times {\bar C}_l^\dagger {\Bar{\Bar {\bf r}}} 
{\bar C}_j + {\bf v}_i \times ({\bar C}_l^\dagger {\Bar{\Bar {\bf r}}} {\bar C}_j) 
({\bar C}_i^\dagger \Bar{\Bar \Delta}^\dagger {\bar C}_l) \}\ , \\
\sum\limits_{l \in \Sigma} \langle \Psi_i |{\bf r}| \Psi_l \rangle \times 
\langle \Psi_l |{\bf r}| \Psi_j \rangle = \sum\limits_{l \in \Sigma} 
{\bar C}_i^\dagger {\Bar{\Bar {\bf r}}} {\bar C}_l \times {\bar C}_l^\dagger 
{\Bar{\Bar {\bf r}}} {\bar C}_j\ .
\end{array}
\end{equation}

\end{widetext}

Substituting these expressions in Eq.~(\ref{Omegaij_r}), we end up with Eq.~(\ref{Curv_r_nonA}).

\begin{acknowledgments}
This work was supported by the International Max Planck Research School for Science and Technology and by the Deutsche Forschungsgemeinschaft (SFB 762). 
We thank Sergey Ostanin who had drown our attention to the papers of N. A. Shilkova and V. P. Shirokovskii. \cite{Shilkova88,Shilkova88_a}
\end{acknowledgments}


\begin{thebibliography}{99}
\bibitem{Berry84} M.~V.~Berry, Proceedings of the Royal Society of London, A 
\textbf{392}, 45 (1984).

\bibitem{Bohm03} \textit{The Geometric Phase in Quantum Systems}, A.~Bohm,
A.~Mostafazadeh, H.~Koizumi, Q.~Niu, and J.~Zwanziger, (Springer Verlag 2003).

\bibitem{Yao04} Y.~Yao, L.~Kleinman, A.H.~MacDonald, J.~Sinova,
T.~Jungwirth, D.~Wang, E.~Wang, and Q.~Niu, Phys. Rev. Lett. \textbf{92},
037204 (2004).

\bibitem{Wang06} X.~Wang, J.~R.~Yates, I.~Souza, and D.~Vanderbilt, Phys.
Rev. B \textbf{74}, 195118 (2006).

\bibitem{Wang07} X.~Wang, D.~Vanderbilt, J.~R.~Yates, and I.~Souza, Phys.
Rev. B \textbf{76}, 195109 (2007).

\bibitem{Guo05} G.~Y.~Guo, Y.~Yao, and Q.~Niu, Phys. Rev. Lett. \textbf{94}, 226601 (2005).

\bibitem{Yao05} Y.~Yao and Z.~Fang, Phys. Rev. Lett. \textbf{95}, 156601 (2005).

\bibitem{Guo08} G.Y.~Guo, S.~Murakami, T.-W.~Chen, and N.~Nagaosa, Phys.
Rev. Lett. \textbf{100}, 096401 (2008).

\bibitem{Haldane04} F.~D.~M.~Haldane, Phys. Rev. Lett. \textbf{93}, 206602
(2004).

\bibitem{Thouless82} D.~J.~Thouless, M.~Kohmoto, M.~P.~Nightingale and M.~den~Nijs, 
Phys. Rev. Lett. \textbf{49}, 405 (1982).

\bibitem{Korringa47} J.~Korringa, Physica \textbf{13}, 392 (1947).

\bibitem{Kohn54} W.~Kohn and N.~Rostoker, Phys. Rev. \textbf{94}, 1111
(1954).

\bibitem{Strange98} {\it Relativistic Quantum Mechanics with
applications in condensed Matter and atomic physics}, P.~Strange, (Cambridge University
Press, 1998).

\bibitem{ZahnPhD} P.~Zahn, Ph.D.~thesis, Technische Universit\"at Dresden, 1998.

\bibitem{Zeller95}  R.~Zeller, P.~H.~Dederichs, B.~\'{U}jfalussy, L.~Szunyogh, P.~Weinberger, Phys. Rev. B \textbf{52}, 8807 (1995).

\bibitem{Elliott54} R.~J.~Elliott, Phys. Rev. \textbf{96}, 266 (1954).

\bibitem{Kramers30} H.~A.~Kramers, Proc. R. Acad. Sci. Amsterdam \textbf{33}, 959 (1930).

\bibitem{Wilczek84} F.~Wilczek and A. Zee, Phys. Rev. Lett. \textbf{52}, 2111 (1984).

\bibitem{Shindou05} R.~Shindou and K.-I.~Imura, Nuclear Physics B \textbf{720}, 399-435 (2005).

\bibitem{Guo08APL} G.~Y.~Guo, J. Appl. Phys. \textbf{105}, 07C701 (2009).

\bibitem{Gradhand09} M.~Gradhand, M.~Czerner, D.~V.~Fedorov, P.~Zahn,
B.~Yu.~Yavorsky, L.~Szunyogh, and I.~Mertig, Phys. Rev. B \textbf{80},
224413 (2009).

\bibitem{Zabloudil2005} {\it Electron Scattering in Solid Matter} J.~Zabloudil, R.~Hammerling, L.~Szunyogh, P.~Weinberger, (Springer Verlag Berlin, 2005)

\bibitem{Shilkova88} N.~A.~Shilkova and V.~P.~Shirokovskii, Phys. Stat. Sol. (b) 
\textbf{149} 571 (1988).

\bibitem{Comment2} In Ref.~\onlinecite{ZahnPhD} the mentioned transformation was
derived for the non-relativistic case. Actually, in the relativistic case it
is similar and was used already by us in Ref.~\onlinecite{Gradhand09}.

\bibitem{Shilkova88_a} N.~A.~Shilkova and V.~P.~Shirokovskii, Phys. Stat. Sol. (b) 
\textbf{149}, 195 (1988).

\bibitem{Guo_95} G.~Y.~Guo and H.~Ebert, Phys. Rev. B \textbf{51}, 12633 (1995).

\bibitem{Resta2000} R.~Resta, J. Phys.: Condens. Matter \textbf{12}, R107 (2000). 

\bibitem{Kalaba81} R.~Kalaba, K.~Spingarn, L.~Tesfatsion, Jour. Optim.
Theor. \& Appl. \textbf{33}, 1 (1981).

\bibitem{Falko} F.~Pientka, Diploma thesis, University Halle-Wittenberg (2010).

\bibitem{Fabian_98} J.~Fabian and S.~D.~Sarma, Phys. Rev. Lett. \textbf{81}, 5624 (1998).

\bibitem{Freimuth_2010} F.~Freimuth, S.~Bl\"ugel, and Y.~Mokrousov, Phys. Rev. Lett. \textbf{105}, 246602 (2010).

\bibitem{Vernes_2007} A.~Vernes, B.~L.~Gyoerffy, and P.~Weinberger, Phys. Rev. B \textbf{76}, 012408 (2007).

\bibitem{Lowitzer_2011} S.~Lowitzer, M.~Gradhand, D. K\"odderitzsch, D.V. Fedorov, I. Mertig, and H. Ebert, Phys. Rev. Lett. \textbf{106}, 056601 (2011).

\bibitem{Lowitzer_2010} S.~Lowitzer, D.~Ködderitzsch, and H. Ebert, Phys. Rev. B \textbf{82}, 140402 (2010)

\bibitem{Sinova_2004} J.~Sinova. D.~Culcer, Q.~Niu, N.~A.~Sinitsyn, T.~Jungwirth, and A. H.~MacDonald, Phys. Rev. Lett. \textbf{92}, 126603 (2004).

\bibitem{Shi2006} J.~Shi, P.~Zhang, D.~Xiao, and Q.~Niu, Phys. Rev. Lett. {\bf 96}, 076604 (2006).

\bibitem{KL_54} R.~Karplus and J.~M.~Luttinger, Phys. Rev. \textbf{95}, 1154
(1954).
\bibitem{Luttinger_58} J.~M.~Luttinger, Phys. Rev. \textbf{112}, 739 (1958).
\bibitem{Sundaram_99} G.~Sundaram and Q.~Niu, Phys. Rev. B \textbf{59}, 14 915 (1999).

\bibitem{Culcer_2004} D.~Culcer, J.~Sinova, N.~A.~Sinitsyn, T.~Jungwirth, A.~H.~MacDonald, and Q.~Niu,  Phys. Rev. Lett. \textbf{93}, 046602 (2004).

\bibitem{Mikitik1999} G.~P.~Mikitik and Yu.~V.~Sharlai, Phys. Rev. Lett. {\bf 82}, 2147 (1999).
\end{thebibliography}
\end{document}